\pdfoutput=1
\documentclass[]{fairmeta}
\usepackage{float}

\usepackage{booktabs}
\usepackage{array}
\usepackage{balance} 
\usepackage{multirow}
\usepackage[normalem]{ulem}
\usepackage{color}
\definecolor{lightgray}{RGB}{215,215,215}
\definecolor{myred}{RGB}{210,109,91}
\usepackage{colortbl}  %彩色表格需要加载的宏包
\usepackage{xcolor}
\useunder{\uline}{\ul}{}
\usepackage{algorithm}  
\usepackage{algorithmicx}  
\usepackage[noend]{algpseudocode}  

\usepackage{amssymb}
\usepackage{amsmath}  
\usepackage{enumitem}
\usepackage{tabularx}
\usepackage[utf8]{inputenc}
\usepackage[english]{babel}
\usepackage{amsthm}
\usepackage{bm}
\newcommand{\ie}{\emph{i.e., }}
\newcommand{\eg}{\emph{e.g., }}

\newcommand{\cf}{\emph{cf. }}

\newcommand{\todo}[1]

\newlength\myindent
\floatname{algorithm}{Algorithm}

\definecolor{darkred}{HTML}{B32B3E}
\definecolor{darkblue}{HTML}{064A6C}

\title{Verifiable Reasoning for LLM-based Generative Recommendation}

\author[]{Xinyu Lin$^1$}
\author[]{Hanqing Zeng$^1$}
\author[]{Hanchao Yu$^1$}
\author[]{Yinglong Xia$^1$}
\author[]{Jiang Zhang$^1$}
\author[]{Aashu Singh$^1$}
\author[]{Fei Liu$^1$}
\author[]{Wenjie Wang$^2$}
\author[]{Fuli Feng$^2$}
\author[]{Tat-Seng Chua$^2$}
\author[]{Qifan Wang$^1$}

\affiliation[]{$^1$Meta Modern Recommendation System (MRS) $^2$National University of Singapore}

\abstract{
Reasoning in Large Language Models (LLMs) has recently shown strong potential in enhancing generative recommendation through deep understanding of complex user preference. 
Existing approaches follow a {reason-then-recommend} paradigm, where LLMs perform step-by-step reasoning before item generation. 
However, this paradigm inevitably suffers from reasoning degradation (\ie homogeneous or error-accumulated reasoning) due to the lack of intermediate verification, thus undermining the recommendation. 
To bridge this gap, we propose a novel \textbf{\textit{reason-verify-recommend}} paradigm, which interleaves reasoning with verification to provide reliable feedback, guiding the reasoning process toward more faithful user preference understanding. 
To enable effective verification, we establish two key principles for verifier design: 
1) \textit{reliability} ensures accurate evaluation of reasoning correctness and informative guidance generation; and
2) \textit{multi-dimensionality} emphasizes comprehensive verification across multi-dimensional user preferences. 
Accordingly, we propose an effective implementation called VRec. 
It employs a mixture of verifiers to ensure multi-dimensionality, while leveraging a proxy prediction objective to pursue reliability. 
Experiments on four real-world datasets demonstrate that VRec substantially enhances recommendation effectiveness and scalability without compromising efficiency. 
}

\date{\today}
\correspondence{Xinyu Lin at \email{xylin1028@gmail.com}, Qifan Wang at \email{wqfcr@meta.com}}
\codes{\url{https://github.com/Linxyhaha/Verifiable-Rec}}

\begin{document}
\maketitle

\section{Introduction}\label{sec:intro}

\begin{figure}[t]
% \vspace{-0.2cm}
\setlength{\abovecaptionskip}{0.02cm}
\setlength{\belowcaptionskip}{-0.3cm}
\centering
\includegraphics[width=0.7\linewidth]{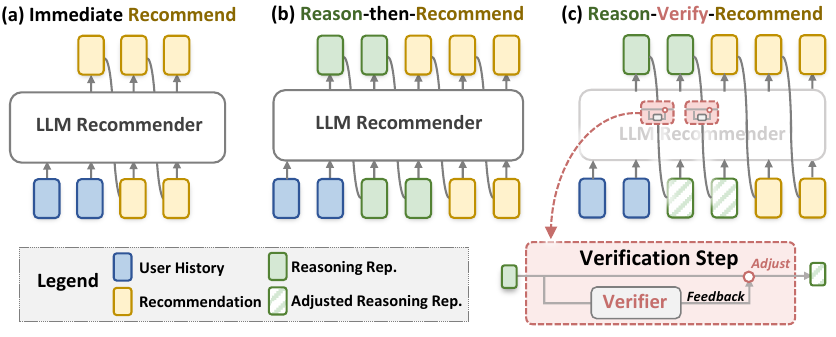}
\caption{Paradigm comparisons between (a) typical LLM-based recommendation, (b) reasoning for LLM-based recommendation, and (c) our proposed verifiable reasoning for LLM-based recommendation.}
\label{fig:paradigm_comparison}
\end{figure}

Incentivizing reasoning in Large Language Models (LLMs) has been extensively explored and achieved remarkable success as exemplified by OpenAI-o1~\citep{openai2025reasoning}, DeepSeek-R1~\citep{guo2025deepseek}, Kimi-K2~\citep{team2025kimi}, and Qwen-QwQ~\citep{team2024qwen2}. 
By performing logical step-by-step reasoning (\eg chain-of-thoughts~\citep{wei2022chain}) prior to producing responses, LLMs substantially enhance their ability 
to tackle complex tasks such as planning~\citep{wei2025plangenllms}, programming~\citep{ding2024reasoning}, and mathematics~\citep{huang2025winninggoldimo2025}. 
In the domain of recommendation, LLM 
reasoning offers the potential to more effectively capture non-trivial and evolving user interests as well as diverse item characteristics~\citep{lin2024bridging}. 
In the light of this, leveraging reasoning for LLM-based generative recommendation (Reason4Rec) has spurred increasing research efforts~\citep{tsai2024leveraging,kim2408review}, emerging as a particularly promising direction for advancing recommender systems.

Technically, given a user’s historical interactions, LLMs aim to generate the next item that aligns with the user’s preference~\citep{bao2025bi}. 
Moving beyond immediate generation that merely capture shallow user preference~\citep{fang2025reason4rec,tang2025think}, Reason4Rec adopts a \textbf{\textit{reason-then-recommend}} paradigm, where LLMs typically first deliberate over user history in the latent space autoregressively\footnote{Other relevant work, specifically designed for rating prediction that reasons in the language space, is discussed in Section~\ref{sec:related_work}.}, and then generate the next recommended item. 
In this paradigm, these autoregressive reasoning steps strive to purify intricate user preferences from the implicit feedback (\ie historical interactions),  
guiding the model towards a deeper understanding that leads to appropriate recommendations. 
Despite the promise, the existing reason-then-generate paradigm faces a fundamental limitation: {unverified reasoning}. 
Without verification, LLMs are prone to two major failure scenarios that lead to reasoning degradation: 
\begin{itemize}[leftmargin=*]
    \item \textit{Homogeneous reasoning}. Lacking supervision on intermediate reasoning steps, the reasoning process might get stuck at the surface level, repeatedly drawing on the same spurious correlations without uncovering new insights~\citep{yuan2024llms,ding2024break} (refer to Section~\ref{sec:preliminaries} for detailed analysis). 
    \item \textit{Error-accumulated reasoning}. In the absence of verification, early missteps in reasoning can propagate and compound across subsequent steps~\citep{zhou2024can}, ultimately amplifying errors and yielding unreliable recommendations (see Figure~\ref{fig:preliminary}(b)). 
\end{itemize}

In light of this, we introduce a novel verifiable reasoning paradigm, which allows LLM to audit the reasoning towards a more faithful user preference understanding based on the verification feedback, thus unlocking the full potential of reasoning. 
Specifically, the verifiable reasoning follows a \textbf{\textit{reason-verify-recommend}} paradigm as shown in Figure~\ref{fig:paradigm_comparison}(c). It contains three key steps:
1) \textbf{Reasoning step} aims to deliberate over the user history to uncover the underlying user preference from implicit feedback. 
Given the intermediate reasoning outcome, 
2) \textbf{verification step} justifies the reasoning outcome and outputs the refined counterpart, which will then be used in the next reasoning step. 
The reasoning and verification steps are executed in an interleaved manner, allowing each verification to guide the subsequent reasoning. 
Lastly, after several reasoning and verification steps, to recommend items, the 
3) \textbf{recommendation step} autoregressively generate the next item based on the reasoning outcome.

Essentially, the verification step bridges between each reasoning step, which evaluates the reasoning and provides guidance signals for adjustment towards more accurate user understanding. 
A critical research question is naturally raised: \textit{How to build an effective verifier}? 
We posit two principles for the verifier design: 
\begin{itemize}[leftmargin=*]
   
    \item \textbf{\textit{Reliability}} 
    ensures that the reasoning output is both properly evaluated and effectively adjusted by the verifier. 
    However, directly assessing the reasoning correctness is challenging, as the reasoning step operates in the latent space without naturally existing supervision to indicate its quality. 
    Therefore, it is essential to design a proxy evaluation task that enables the verifier to provide feedback that can approximate the quality of each reasoning step. 
    Moreover, beyond providing evaluative feedback, the verifier should generate guidance signals that can effectively revise the intermediate reasoning representations, thereby mitigating degradation from homogeneous and error-accumulated reasoning. 
    \item \textit{\textbf{Multi-dimensionality}} requires the verifier to justify the intermediate reasoning outcome from multiple perspectives. 
    Relying on a single aspect (\eg item category) is insufficient to capture alignment with rich user preferences. 
    Hence, we consider two complementary forms of dimension diversity. 
    \textit{Intra-user diversity} justify the multi-aspect user preference alignment across different aspects (\eg semantic and collaborative information)
    ; Meanwhile, 
    \textit{inter-user} diversity emphasizes that different users might focus on different aspects of item (\eg a user might like the song ``My Heart Will Go On'' because of the associated movie Titanic while another user might like it due to the singer). 
\end{itemize}

\begin{figure}[t]
% \vspace{-0.2cm}
\setlength{\abovecaptionskip}{0.02cm}
\setlength{\belowcaptionskip}{-0.3cm}
\centering
\includegraphics[scale=0.7]{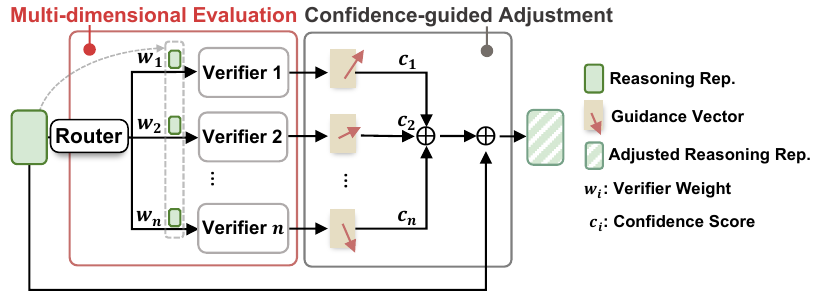}
\caption{Illustration of verification step in ``reason-verify-recommend'' paradigm. The multi-aspect verification ensures the verification diversity, while confidence-based adjustment emphasizes verification certainty.}
\label{fig:overview}
\end{figure}

Based on the above principles, we propose VRec, an effective implementation of the verifiable reasoning paradigm for LLM-based {Rec}ommendation. 
To ensure the verification multi-dimensionality, we design a novel mixture of verifiers as shown in Figure~\ref{fig:overview}. 
Specifically, for intra-user dimension diversity, we employ a set of verifiers, each dedicated to a specific aspect such as item category, title semantics, and collaborative information. 
For inter-user dimension diversity, we further incorporate a personalized router to adaptively weights the contributions of different verifiers according to individual user behaviors. 
On the other hand, to ensure the verification reliability, we design a group-level preference prediction task as a proxy objective for verifier, where we leverage the verifier's internal signals, \ie prediction entropy and model weights, as evaluation feedback and the guidance signals, respectively, for reasoning adjustment. 
To instantiate VRec, we train the verifiers and the LLM recommender via a two-stage training strategy, where we introduce a novel monotonicity regularization to encourage progressively more accurate reasoning. 
Extensive experiments on four real-world datasets validate the effectiveness, efficiency, and promising scalability of the proposed method.

The main contributions of this work are summarized as follows:
\begin{itemize}[leftmargin=*]
    \item We highlight the importance of verification in LLM reasoning and introduce a novel reason-verify-recommend paradigm for LLM-based generative recommendation, with two principles for verifier design. 
    \item To instantiate the paradigm, we propose an effective implementation VRec, which employs a mixture of verifiers to capture diverse user preference and deliver reliable feedback on intermedate reasoning steps, exploiting the potential of LLM reasoning. 
    \item We conduct extensive experiments on four real-world datasets to validate the practicality of the verifiable reasoning paradigm, and justify the effectiveness, scalability, and diversity of VRec. 
\end{itemize}

\section{Task Formulation}\label{sec:preliminaries}
In this section, we first introduce the existing \textit{reason-then-recommend} paradigm of Reason4Rec, and uncover the issues of this paradigm. Then, we formulate the paradigm of verifiable reasoning. 

\vspace{2pt} 
\noindent$\bullet\quad${\textbf{Retrospect of Reason4Rec.}} 
Given a user's historical interactions $X=[i_1, i_2, \dots, i_L]$ in chronological order, where $i \in\mathcal{I}$ is the item identifier (\eg item title) and $L=|X|$, LLM-based recommendation aims to generate the next item $y=i_{L+1}$ that matches the user preference. 

\vspace{2pt}
\textit{\textbf{Reason-then-recommend paradigm}}. 
To leverage reasoning for recommendation, prior studies mainly employ reasoning before generating the next item. 
Formally, given input $\bm{x}$, an LLM  $\mathcal{M}$ parameterized by $\theta$ first generates a reasoning sequence:
\begin{equation}\label{eqn:reaoning_step}
\begin{aligned}
    {R} &= (\bm{r_1}, \bm{r_2}, \dots, \bm{r}_m), \quad \text{where } 
    \bm{r}_t = \mathcal{M}(\cdot \mid X, \bm{r}_{1:{t-1}}; \theta), 
\end{aligned}
\end{equation} 
where each intermediate reasoning representation $\bm{r}_j\in \mathbb{R}^{d_{\mathcal{M}}}$ is a latent representation~\citep{zhang2025reinforced} with LLM's hidden dimension $d_{\mathcal{M}}$. 
Conditioned on both the user history $X$ and reasoning $R$,  
the LLM then generates the next item:
\begin{equation}\label{eqn:recommend_step}
\begin{aligned}
    \hat{y}_t = \arg\max_{v \in \mathcal{V}} \mathcal{M}(v \mid X, R, \hat{y}_{<t}; \theta),
\end{aligned}
\end{equation} 
where $\mathcal{V}$ is the item vocabulary, $\hat{y}_t$ is the $t$-th token of $\hat{y}$, and $\hat{y}_{<t}$ represents the token sequence preceding $y_t$. 
This paradigm introduces additional step-by-step reasoning, thus augmenting user behavior understanding and leading to better recommendations. 

\begin{figure}[t]
\vspace{-0.2cm}
\setlength{\abovecaptionskip}{0.02cm}
\setlength{\belowcaptionskip}{-0.3cm}
\centering
\includegraphics[scale=0.8]{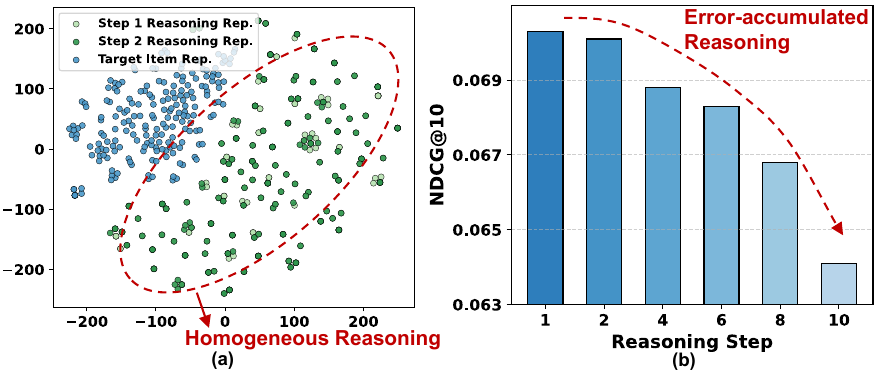}
\caption{(a) illustrates the t-SNE visualization of the representations of the latent reasoning and the target item of LatentR$^3$ with 2 reasoning steps. (b) demonstrates continuous performance degradation as the reasoning step increases under the reason-then-recommend paradigm.}
\label{fig:preliminary}
\end{figure}

\vspace{2pt}
\textit{\textbf{Issues of the paradigm.}}  
While promising, the reasoning process is \textit{unverified} due to the lack of feedback on intermediate reasoning outcomes. 
This can lead to two consequences: 
\begin{itemize}[leftmargin=*]
    \item \textit{{Homogeneous reasoning}}. Without intermediate verification, LLMs might shortcut the reasoning process and exploit spurious correlations. 
    Without loss of generality, LLM recommender is optimized via: 
    \begin{equation}\small
    \label{eqn:llm_opt}
    \begin{aligned}
        \mathop{\arg\min}_{\theta\in\Theta} \space \mathbb{E}_{X \sim \mathcal{D}} \mathbb{E}_{R\sim \mathcal{M}} \mathcal{L}_\theta( \hat{{y}}\mid X, R),
    \end{aligned}
    \end{equation}
    where $\mathcal{L}_\theta(\cdot)$ can be substituted by the form of training objective in supervised fine-tuning (\eg generative loss~\citep{zhu2024collaborative}) or reinforcement learning (\eg policy-gradient objective~\citep{you2025text}). 
    However, the supervision signal is only provided on the recommended item $\hat{y}$, neglecting the direct optimization of $R$. 
    As such, the model may converge to a degenerate distribution that collapses to trivial or homogeneous reasoning patterns (see Appendix~\ref{app:detailed_descriptions} for detailed explanations). 
    To empirically validate this issue, we visualize the embedding of the intermediate reasoning as shown in Figure~\ref{fig:preliminary}(a). 

    \item \textit{{Error-accumulated reasoning}}. In the absence of verification, the autoregressively generated reasoning representation will suffer from the error accumulation issue. In other words, the initial inaccurate reasoning will eventually lead to inappropriate recommendations. 
    Therefore, the LLM recommender struggles to achieve deep reasoning, hindering the full utilization of reasoning capability (empirical evidence in Figure~\ref{fig:preliminary}(b)). 
\end{itemize}
 
\vspace{2pt}
\noindent$\bullet\quad$\textbf{Verifiable Reasoning for LLM-based Recommendation.} 
To overcome the limitations, we introduce a novel {reason-verify-recommend} paradigm, which incorporates an additional verification step for reasoning representations. Formally, for any intermediate reasoning representation $\bm{r}\sim \mathcal{M}(X,R_{<t})$ at step $t$, we have
\begin{equation}\label{eqn:verifiable_paradigm}
\begin{aligned}
\begin{cases}
    f, g \leftarrow \text{Verifier}(\bm{r}), \\
    \bm{r}^{*} \leftarrow \text{Adjuster}(\bm{r}, f, g),
\end{cases}
\end{aligned}    
\end{equation}
where $f$ and $g$ is the evaluation feedback and the guidance signals respectively, and $\bm{r}^{*}$ is the adjusted reasoning representation, which will then substitute the initial $\bm{r}$ for the next reasoning step, \ie $R_{\leq t}=(\bm{r}_1^{*},\dots, \bm{r}_{t}^{*})$. 
After interleaving reasoning and verification for $m$ steps, the item is generated via Eq.(\ref{eqn:recommend_step}). 

The key to an effective verifiable reasoning lies in the design of the verifier, for which we posit two fundamental principles: 1) multi-dimensionality aims to evaluate $\bm{r}$ from multiple aspects to ensure preference diversity within users and across users. 2) Reliability emphasizes a proper design of evaluation feedback $f$ and guidance $g$, to effectively reflect reasoning quality and guiding direction, thus achieving reliable reasoning adjustment.

\section{VRec}\label{sec:method}
\begin{figure*}[t]
% \vspace{-0.2cm}
\setlength{\abovecaptionskip}{0.02cm}
\setlength{\belowcaptionskip}{-0cm}
\centering
\includegraphics[scale=0.7]{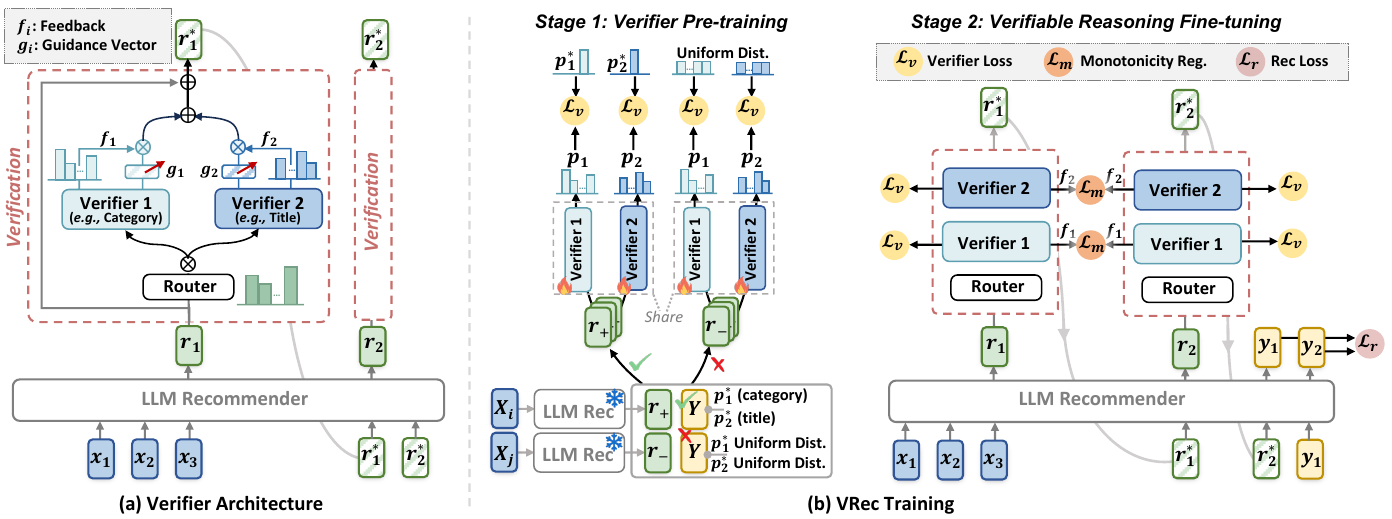}
\caption{Overview of VRec. (a) shows the design of the mixture of verifiers, including a personalized router, and a set of verifiers for different dimensions. We demonstrate two verifiers as an example for demonstration clarity. (b) shows the two-stage training strategy of VRec, including verifier pre-training and verifiable reasoning fine-tuning.}
\label{fig:method}
\end{figure*}

To pursue the two objectives, we propose an effective implementation VRec. As illustrated in Figure~\ref{fig:method}, VRec employs a set of verifier models to evaluate the reasoning outcome, along with the feedback $f$ and the guidance signals (Section~\ref{sec:verifier_architecture}). 
To train the verifier, we propose a two-stage training strategy, including verifier pre-training, and verifiable reasoning fine-tuning (Section~\ref{sec:training}).

% overview, introducing the pipeline of the reason-verify-recommend workflow. The 

\subsection{Verifier Architecture}\label{sec:verifier_architecture}
% objective: intra-user verification diversity, and inter-user verification diversity. 

% motivation: 首先要说明我们希望在abstract-level上（就是group level)对reasoning做出一个verification
To unlock the potential of the verifiable reasoning paradigm, we aim to design a verifier with two key considerations \ie multi-dimensionality and reliability. The architecture is illustrated in Figure~\ref{fig:method}(a). 

\vspace{2pt}
\noindent$\bullet\quad$\textbf{Mixture of Verifiers}. 
To achieve \textit{intra-user dimension diversity}, 
\ie user's preference can be multi-dimensional, we employ a set of verifiers $\mathcal{V} = \{V_1, V_2, \dots, V_{n}\}$, where each verifier $V_i$ focuses on a specific aspect, such as category and collaborative information. 
% This encourages a more comprehensive verification of the reasoning quality. 
For each verifier $V_i$, it evaluates the alignment between the intermediate reasoning representation and the user preference over the specific aspect $i\in\{1,\dots, n\}$: 
\begin{equation}
f_i, g_i \leftarrow V_i(\bm{r}),
\end{equation}
where $f_i$ and $g_i$ is the evaluation feedback and guidance for the aspect $i$. The set of feedback and guidance is then denoted as $\mathcal{F}=\{f_1, \dots, f_n\}$ and $\mathcal{G}=\{g_1, \dots, g_n\}$, respectively, which will then be used for reasoning adjustment. 

\textit{\textbf{Personalized router}}. In addition to the intra-user dimension diversity, we also consider \textit{inter-user dimension diversity}, \ie different users may prioritize different aspects. For instance, while some users may focus more on the item category (\eg jazz music), some users may be more concerned with the price. 
Therefore, we introduce a personalized router for the set of verifiers to obtain the personalized verification weights for each aspect: 
\begin{equation}
\begin{cases}
    \bm{w} = g(\bm{r_i}), \quad \text{where }\bm{w}=[w_1, \dots, w_n] \in \mathbb
R^n \\
    f_i, g_i \leftarrow V_i(w_i\bm{r}),
\end{cases}
\end{equation}
where $g(\cdot)$ is the personalized router to assign adaptive weights for different verification aspects according to individual user behaviors. 
% 首先要讲清楚，最直接的方式是训练一个分类器知道是对是错。但是这个过程很难收集。alternatively, 我们选择一个
% To verify the intermediate reasoning outcome $r$, an intuitive solution is to design the verifier $V$ to do a binary classification on the correctness of the reasoning outcome. 
% Nonetheless, it is difficult to obtain such supervision signals as 

\vspace{2pt}
\noindent$\bullet\quad$\textbf{Design of Evaluation Feedback ${f_i}$}. 
% In VRec, we leverage the entropy and the hidden states of the verifier to present $F_i$ and $G_i$, respectively. 
To verify the intermediate reasoning outcome $\bm{r}$, an intuitive solution is to consider the verification task as a binary classification on the correctness of the reasoning outcome, \ie $f_i$ is presented as 0 or 1. 
Nonetheless, training such a verifier is implausible since it is difficult to obtain the ground-truth reasoning correctness from implicit user feedback (\eg clicks) in recommendation data. 
Worse still, the latent reasoning process that leads to accurate recommendations can vary in granularity and dimensionality, introducing substantial noise that hampers building an effective binary classifier. 

\textbf{\textit{Preference prediction task}}. 
Alternatively, reasoning that leads to accurate recommendations should align with the group-level
user preference (\ie user's coarse-grained interests such as hip-hop music). 
In this sense, rather than evaluating reasoning at arbitrary levels of granularity, we focus on identifying and refining the reasoning process that is likely to conflict with the user’s broad preference. 
Therefore, we define a preference prediction task for the verifier as a proxy of binary evaluation: 
% \HQ{``group'' sounds a bit vague. Are we saying that throughout reasoning process, the model can gradually refine the prediction. So ``group'' means from coarse to fine grained prediction?} 
% In other words, while we do not expect to justify preference reasoning of arbitrary granularity, we seek to revise the reasoning that is highly likely to conflict with user's broad preference.
\begin{equation}\small\label{eqn:preference_prediction}
\begin{cases}
    \bm{p} = V_i(w_i\bm{r}), \quad\text{where } \bm{p}=[p_1,\dots,p_{d_i}],\\
    \hat{P} = \mathop{\arg\max}_{j} p_j, 
\end{cases}
\end{equation}
where $\bm{p}\in\mathbb{R}^{d_i}$ is the probability distribution of the preference prediction, $d_i$ is a pre-defined class number of the group-level preference, and $\hat{P}$ is the predicted group-level preference (\eg ``hip-hop'' for category). 
Accordingly, the verifier $V_i$ is a learnable classifier, which can be implemented by a linear head or an MLP (detailed discussions of the classifier design are in Section~\ref{sec:verifier_size_scalability}). 

\textbf{\textit{Entropy-based evaluation}}. 
Intuitively, with a well-trained verifier\footnote{Refer to Section~\ref{sec:training} for verifier training details.}, the reasoning representation $\bm{r}$ is aligned with the group-level user preference $P^{*}$ if $\hat{P}=P^{*}$, otherwise misaligned. 
However, the $P^{*}$ is unknown for real-time recommendation, we thus utilize the entropy to estimate the preference alignment as:
\begin{equation}\small\label{Eq:f}
    f_i = H(\bm{p}) = - \sum_{j=1}^{d_i} p_j \log p_j. 
\end{equation}
The entropy typically indicates the model certainty and is widely used to measure the level of confidence~\citep{fu2025deep,farquhar2024detecting,liu2024uncertainty}. 
As such, a small $f_i$ implies that the reasoning is aligned with the preference $P^{*}$ of a high confidence; 
while a large $f_i$ means that the reasoning is less confident and might conflict against preference $P^{*}$. 

% To predict the preference, we implement the verifier $V_i$ as a linear layer or an MLP. Detailed analysis of verifier architecture can be found in Section~\ref{}. 

% classification task for the verifier, which aims to predict the high-level preference over the specific dimension (\eg preference on ``hiphop'' from the category dimension). 
% Precisely, the verifier evaluates the reasoning outcome 

% 所以我们引入了high-level category
% verifier的output是多分类任务，我们利用已知的
% for intra-user diversity
% input: intermediate reasoning steps. 
% output: categorical prediction. 

% 这个category prediction 是一个verification
% We regard the verifier’s categorical prediction as a proxy for reasoning reliability. Specifically, a confident prediction (low entropy distribution) suggests that the intermediate reasoning outcome aligns well with a coherent preference signal, whereas a high-entropy distribution indicates ambiguity, thereby questioning the validity of the reasoning path. In this way, the verifier serves as a principled mechanism to audit and validate the reasoning process.

\vspace{2pt}
\noindent$\bullet\quad$\textbf{Design of Guidance Signal ${g_i}$}. 
To adjust the intermediate reasoning outcome, we then need to obtain the guidance signal $\bm{g}_i$ to adjust the reasoning accordingly, thus preventing the homogeneous and error-accumulated reasoning issues. 
Specifically, prior studies reveal that output predictions are anchored by embedding vectors that act as prototypes~\citep{morin2005hierarchical,li2018deep,zhou2022prototype}. Inspired by this, we utilize the weights of the last layer of the verifier as preference prototypes for reasoning adjustment. We obtain 
\begin{equation}\small
    \bm{g}_i = \bm{W}^{i}_{:, j^{*}}, \quad j^{*} = \mathop{\arg\max_j p_j}, 
\end{equation}
where $\bm{W}^{i}\in\mathbb{R}^{d_\mathcal{M}\times d_i}$ is the weights of the verifier's last layer, and
$d_\mathcal{M}$ is the hidden dimension of the LLM.

\textbf{\textit{Confidence-based adjustment}}. 
The guidance vector $\bm{g}_i\in\mathbb{R}^{d_\mathcal{M}}$ then adjusts the reasoning representation via: 
\begin{equation}\small
\label{Eq:adjustment}
    \bm{r}^{*} = \sum_{i=1}^{n} (1-c_i)\bm{r} + c_i\bm{g}_i,\quad\text{where }c_i=\frac{1}{f_i} \text{ is the confidence score}.
\end{equation} 
Intuitively, for the reasoning that has a good alignment with group-level user preference, \ie a large $f_i$, we seek to strengthen the reasoning over the current path, aggressively boosting the reasoning, thus alleviating the homogeneous issue; 
On the contrary, if the reasoning diverges from the user preference $P^{*}$, we aim to prevent the propagation of the error and encourage the model to explore other reasoning paths with a modest combination of initial reasoning outcome and the guidance vector. 
% \todo{revise the definition}

\subsection{VRec Training}\label{sec:training} 
The VRec architecture comprises two learnable components, \ie the verifiers and the LLM recommender. To ensure reliable feedback from the verifier and seamless integration with the recommender, we adopt a two-stage training strategy as shown in Figure~\ref{fig:method}(b).

\vspace{2pt}
\noindent$\bullet\quad$\textbf{Stage 1: Verifier Pre-training}. 
The goal of verifier pre-training is to train a verifier that can correctly predict the group-level preference a reliable reasoning representation should capture. 
By teaching the verifier to give good prediction on group-level preference, we can obtain reliable evaluation feedback $f$ and informative guidance $\bm{g}$ for effective adjustment. 

\textbf{\textit{Pre-training data collection.}} To train the verifier, we first collect the pre-training data. 
Specifically, we utilize a pre-trained LLM reasoning recommender to generate the reasoning and corresponding recommendations for all samples $(X,y)\in\mathcal{D}$. 
For each sample, if the target next item is successfully generated, we consider it as a positive sample and collect the reasoning outcome $R$ and pair it with the corresponding preference $P$ for each aspect of the target item (\eg ``jazz'' for category aspect)\footnote{Refer to Appendix~\ref{app:detailed_descriptions} for detailed implementation of group-level preference labeling.}. 
Conversely, for the samples that fail to generate the target next item, \ie negative samples, we pair the reasoning outcome $R$ with $\emptyset$.
The obtained dataset is denoted as ${\mathcal{D}_v}=\{(R,\mathcal{P})\mid \mathcal{P}=\{P_1, \dots,P_n\} \text{ or } \emptyset \}$.

\textbf{\textit{Training objective.}} 
The verifier is then optimized via minimizing the cross entropy on positive samples while maximizing the entropy on negative samples. We define the loss function as:
\begin{equation}\small
\label{Eq:verifier_loss}
    \mathcal{L}_v=
    \begin{cases}
        -\log \text{Pr}(\hat{P}=P), \quad \text{if } P\neq \emptyset, \\
        - \alpha\sum_{j=1}^{d_i} p_j \log p_j, \quad \text{if } P=\emptyset,
    \end{cases}
\end{equation}
where $\hat{P}$ is the predicted preference via Eq.(\ref{eqn:preference_prediction}), and $\alpha$ is a hyper-parameter that denotes the weight of the negative samples. 
Intuitively, we encourage the verifier to predict accurately on a reasoning that is preference-aligned. 
On the contrary, for the low-quality reasoning, we encourage the verifier to output a uniform distribution, \ie a high-entropy prediction that lead to a large $f$. 

\vspace{2pt}
\noindent$\bullet\quad$\textbf{Stage2: Verifiable Reasoning Fine-tuning}. 
Subsequently, based on the pre-trained verifier and LLM recommender, we further jointly fine-tune the two models in an end-to-end manner, enabling verifiable reasoning that can ultimately lead to appropriate recommendations. Specifically, we have 
\begin{equation}\small\label{Eq:llm_loss}
    \mathcal{L}_{r} = -\frac{1}{|\mathcal{D}|}\sum_{\mathcal{D}}\sum_{t=1}^{|y|}\mathcal{M}_\theta(X, R, y_{<t}),
\end{equation}
where $R=(\bm{r}_1^{*}, \dots, \bm{r}_2^{*})$, and each $r^{*}$ is obtained via Eq.(\ref{Eq:adjustment}). 

\textbf{\textit{Monotonicity regularization}}. 
% It is noted that sharing a verifier across different steps might also find shortcut and hinders the deep reasoning. 
It is worth noting that sharing the verifiers across multiple verification steps may also suffer from shortcut learning~\citep{geirhos2020shortcut}, thereby hindering deeper reasoning capability. 
To address this issue, we propose a monotonicity regularization to further enforce progressively more accurate reasoning, which is defined as:
\begin{equation}\small\label{eqn:monotonicity_reg}
    \mathcal{L}_m = \max(0, f^t_i - f^{t-1}_i),
\end{equation}
where $f^t_i = H(\bm{p})$ is the evaluation feedback at $t$-th reasoning step as defined in Eq.(\ref{Eq:f}). 
By penalizing increases in entropy between consecutive steps, this regularization encourages the reasoning process to become increasingly aligned with the group-level user preference. 
 
\textbf{\textit{Overall loss}}. 
To ensure the verifier's capability of evaluating the alignment between the reasoning and the user preference, we incorporate the verifier loss (Eq.(\ref{Eq:verifier_loss})) as a regularization. 
To summarize, the overall loss for the Stage 2 is:
\begin{equation}\small
\label{Eq:stage3_loss}
    \mathcal{L} = \mathcal{L}_r + \beta \mathcal{L}_v + \gamma\mathcal{L}_m,
\end{equation}
where $\beta$ and $\gamma$ are the hyper-parameters to balance the verifier loss and monotonicity regularization, respectively. 

\subsection{VRec Instantiation} 
To instantiate VRec on a pre-trained reasoning LLM recommender with reasoning step $m$, we first pre-train the verifier to evaluate the reasoning via Eq.(\ref{Eq:verifier_loss}). 
Subsequently, we jointly train the verifier and LLM recommender to achieve recommendation-oriented verifiable reasoning via Eq.(\ref{Eq:stage3_loss}). 
For inference, VRec justifies and adjusts each intermediate reasoning outcome via Eq.(\ref{Eq:adjustment}). After performing interleaved reasoning and verification steps, the LLM recommender generates the next item via Eq.(\ref{eqn:recommend_step}). 
The detailed VRec instantiation process is illustrated in Algorithm~\ref{algo:VRec} in the Appendix. 

\vspace{2pt}
\noindent$\bullet\quad$\textbf{Computational Overhead Analysis}. 
In our verifiable reasoning paradigm, the overall complexity consists of two parts. 
Considering multiply–add operations (FLOPs) per forward pass, 
the LLM-based recommender dominates with a complexity of $\mathcal{O}(mL d^2)$, where $m$ is the number of reasoning steps, $L$ the number of transformer layers, and $d$ the hidden dimension. 
The verification module adds $\mathcal{O}(mnkd^2)$ complexity, as each step passes the reasoning embedding through a personalized router and $n$ verifiers, each a $k$-layer MLP.
Despite this addition, the verifier remains lightweight, since its computation focuses on a small number of low-dimensional preference classes (around 20), and its size is negligible compared to the LLM backbone. 
The time costs analysis is also empirically studied in Section~\ref{sec:efficiency_analysis}.

\section{Experiments}\label{sec:exp}
In this section, we conduct experiments to answer the following research questions: 
\textbf{RQ1}: How does the VRec perform compared to the baselines? 
\textbf{RQ2}: How do the different components of VRec (\ie multi-dimensional verifiers, personalized router, monotonicity regularization, and verifier pre-training) affect the performance? 
\textbf{RQ3}: How does VRec perform when scaling up the reasoning steps and verifiers? 
\textbf{RQ4}: How is the computational overhead of the verification step, and how is the hyper-parameter sensitivity? 

\subsection{Experimental Settings}

\subsubsection{\textbf{Datasets}}
We conduct experiments on four real-world datasets from diverse domains, including music, micro-videos, and books. 
Specifically, we adopt two widely used Amazon review datasets\footnote{\url{https://jmcauley.ucsd.edu/data/amazon/}.}, 
1) \textbf{CDs} and 2) \textbf{Instruments}, 
which cover user interactions on music albums and instrument products with rich textual meta information such as titles and categories. 
In addition, we use a micro-video dataset 3) \textbf{MicroLens}\footnote{\url{https://github.com/westlake-repl/MicroLens/.}}, which contains extensive user interactions and textual descriptions for micro-videos~\citep{ni2023content}. 
Furthermore, we adopt 4) \textbf{Goodreads}\footnote{\url{https://cseweb.ucsd.edu/~jmcauley/datasets/goodreads.html/.}}, a book recommendation dataset that provides user interactions on books, where each item is provided with a book title, descriptions, and categories. 
For each dataset, we sort user interactions chronologically based on timestamps and split them into training, validation, and testing sets with a ratio of 8:1:1. 
For each training instance, following~\citep{bao2024decoding}, we constrain the maximum length of item sequence to 10. 
The statistics of the four datasets are summarized in Table~\ref{tab:dataset_statistics} of Appendix. 

\noindent$\bullet\quad$\textbf{Evaluation.} 
To evaluate the models, we adopt the widely used Recall@$K$ (R@$K$) and NDCG@$K$ (N@$K$)~\citep{wang2021denoising}, with $K=5$ and $10$ to assess the recommendation accuracy. 
% In addition, we test the average inference time costs (s) per sample to analyze the model efficiency. 

\begin{table*}[t]
\setlength{\abovecaptionskip}{0.05cm}
\setlength{\belowcaptionskip}{0.2cm}
\caption{Overall performance of VRec and baselines on four real-world datasets. The best results are in bold and the second-best results are underlined. $*$ implies the improvements of VRec over the best baseline results are statistically significant ($p$-value < 0.01) under one-sample t-tests.}
\setlength{\tabcolsep}{2mm}{
\resizebox{\textwidth}{!}{
\begin{tabular}{l|cccc|cccc|cccc|cccc}
\toprule
 & \multicolumn{4}{c|}{\textbf{CDs}} & \multicolumn{4}{c|}{\textbf{MicroLens}} & \multicolumn{4}{c|}{\textbf{Goodreads}} & \multicolumn{4}{c}{\textbf{Instruments}} \\ \midrule
\textbf{Method} & \textbf{R@5} & \textbf{R@10} & \textbf{N@5} & \textbf{N@10} & \textbf{R@5} & \textbf{R@10} & \textbf{N@5} & \textbf{N@10} & \textbf{R@5} & \textbf{R@10} & \textbf{N@5} & \textbf{N@10} & \textbf{R@5} & \textbf{R@10} & \textbf{N@5} & \textbf{N@10} \\ \hline
\textbf{GRU4Rec} & 0.0481 & 0.0671 & 0.0365 & 0.0426 & 0.0036 & 0.0068 & 0.0021 & 0.0031 & 0.0190 & 0.0332 & 0.0117 & 0.0163 & 0.0766 & 0.0938 & 0.0553 & 0.0609 \\
\textbf{Caser} & 0.0549 & 0.0749 & 0.0403 & 0.0467 & 0.0036 & 0.0063 & 0.0020 & 0.0029 & 0.0214 & 0.0378 & 0.0131 & 0.0183 & 0.0775 & 0.0972 & 0.0544 & 0.0608 \\
\textbf{SASRec} & 0.0784 & 0.0984 & 0.0576 & 0.0641 & 0.0044 & 0.0069 & 0.0029 & 0.0037 & 0.0182 & 0.0316 & 0.0114 & 0.0157 & 0.0825 & 0.0972 & 0.0726 & 0.0772 \\ \midrule
\textbf{TIGER} & 0.0381 & 0.0555 & 0.0269 & 0.0325 & 0.0052 & 0.0075 & 0.0023 & 0.0030 & 0.0202 & 0.0384 & 0.0116 & 0.0175 & 0.0760 & 0.0945 & 0.0557 & 0.0616 \\
\textbf{LETTER} & 0.0417 & 0.0619 & 0.0285 & 0.0350 & 0.0053 & 0.0104 & 0.0028 & 0.0045 & 0.0215 & 0.0377 & 0.0134 & 0.0185 & 0.0707 & 0.1012 & 0.0483 & 0.0583 \\
\textbf{SETRec} & 0.0627 & 0.0942 & 0.0454 & 0.0553 & 0.0094 & 0.0129 & 0.0055 & 0.0066 & 0.0106 & 0.0174 & 0.0066 & 0.0088 & 0.0786 & 0.0864 & 0.0721 & 0.0746 \\
\textbf{D3} & 0.0747 & 0.0940 & 0.0596 & 0.0659 & 0.0076 & 0.0134 & 0.0057 & 0.0076 & 0.0328 & 0.0534 & 0.0218 & 0.0285 & 0.0996 & 0.1188 & 0.0853 & 0.0915 \\
\textbf{LatentTTS} & 0.0730 & 0.0912 & 0.0578 & 0.0643 & 0.0069 & 0.0123 & 0.0050 & 0.0068 & 0.0312 & 0.0513 & 0.0206 & 0.0265 & 0.0911 & 0.1068 & 0.0801 & 0.0866 \\ 
\textbf{LatentR$^3$} & 0.0801 & 0.0978 & 0.0645 & 0.0703 & 0.0089 & {\ul 0.0160} & 0.0065 & 0.0088 & 0.0349 & 0.0565 & 0.0231 & 0.0301 & 0.1022 & 0.1243 & 0.0881 & 0.0951 \\ \midrule
\cellcolor{gray!8}\textbf{VRec-1Step} & \cellcolor{gray!8}{\ul 0.0836} & \cellcolor{gray!8}{\ul 0.1070} & \cellcolor{gray!8}{\ul 0.0655} & \cellcolor{gray!8}{\ul 0.0730} & \cellcolor{gray!8}{\ul 0.0100} & \cellcolor{gray!8}0.0154 & \cellcolor{gray!8}{\ul 0.0076} & \cellcolor{gray!8}{\ul 0.0093} & \cellcolor{gray!8}{{\ul 0.0365}} & \cellcolor{gray!8}\textbf{{0.0593}} & \cellcolor{gray!8}{{\ul 0.0236}} & \cellcolor{gray!8}{{\ul 0.0310}} & \cellcolor{gray!8}{{\ul 0.1067}} & \cellcolor{gray!8}{\ul 0.1269} & \cellcolor{gray!8}{\ul 0.0911} & \cellcolor{gray!8}{\ul 0.0976} \\
\cellcolor{gray!18}\textbf{VRec} & \cellcolor{gray!18}\textbf{0.0963*} & \cellcolor{gray!18}\textbf{0.1213*} & \cellcolor{gray!18}\textbf{0.0782*} & \cellcolor{gray!18}\textbf{0.0862*} & \cellcolor{gray!18}\textbf{0.0121*} & \cellcolor{gray!18}\textbf{0.0235*} & \cellcolor{gray!18}\textbf{0.0085*} & 
\cellcolor{gray!18}\textbf{\textbf{0.0121*}} & \cellcolor{gray!18}\textbf{\textbf{0.0366*}} & 
\cellcolor{gray!18}{{\ul 0.0582*}} &
\cellcolor{gray!18}\textbf{\textbf{0.0242*}} &
\cellcolor{gray!18}\textbf{\textbf{0.0312*}} &
\cellcolor{gray!18}\textbf{0.1098*} &
\cellcolor{gray!18}\textbf{0.1301*} &
\cellcolor{gray!18}\textbf{0.0944*} &
\cellcolor{gray!18}\textbf{0.0997*} \\ \bottomrule
\end{tabular}
}}

\label{tab:overall_performance}
\end{table*}

\subsubsection{\textbf{Baselines}}
We compare VRec with various competitive baselines, including traditional discriminative models, LLM-based generative models, and the LLM reasoning-enhanced models. Specifically, 
1) \textbf{GRU4Rec}~\citep{hidasi2015session} employs the gated recurrent unit (GRU) to capture the item's sequential patterns. 
2) \textbf{Caser}~\citep{tang2018personalized} utilizes convolutional neural networks (CNNs) to capture both sequential patterns and general user preference. 
3) \textbf{SASRec}~\citep{kang2018self} is one of the most representative sequential model, which leverages the bi-directional self-attention mechanism to effectively model the complex user behavior. 
4) \textbf{TIGER} ~\citep{rajput2023recommender} is a representative generative recommendation method that employs an RQ-VAE with codebooks to represent each item with a sequence of external tokens. 
5) \textbf{LETTER}~\citep{wang2024learnable} is also a external-token-based method, which integrates both semantic and collaborative filtering signals into item identifier to achieve recommendation that captures the collaborative similarity of items. 
6) \textbf{SETRec}~\citep{lin2025order} is a recent method that introduces order-agnostic item identifiers to better model user preference across different dimensions. 
7) \textbf{D3}~\citep{bao2024decoding} utilizes the item title with human vocabulary to represent each item, and emphasizes the more important tokens with collaborative information to achieve reliable recommendations. 
8) \textbf{LatentR$^3$}~\citep{zhang2025reinforced} is a recent reasoning-enhanced LLM-based method, where LLM first reasons the user preference in the latent space via an additional attention module before generating the next item. 
We also extend a parallel latent reasoning method, 9) \textbf{LatentTTS}~\citep{you2025parallel}, which leverages Monte Carlo dropout and Additive Gaussian noise to diversify the latent reasoning and select paths with high confidence. 
For other existing trustworthy LLM reasoning methods that could not be directly extended to recommendation, we provide detailed explanations in Appendix~\ref{app:detailed_descriptions}).

\subsubsection{\textbf{Implementation Details.}} 
For traditional methods, we search the learning rate and weight decay in the range of $\{1e^{-3}, 1e^{-4}, 1e^{-5}\}$ and $\{1e^{-3},1e^{-4}, 1e^{-5}, 1e^{-6}\}$, respectively. 
For both LLM-based baselines and VRec, we instantiate them on Qwen2.5-1.5B~\citep{qwen2025qwen25technicalreport} and fine-tune the full LLM parameters with learning rate searched from $3e^{-4},5e^{-5}, 3e^{-5}$. 
For LatentR$^3$ and VRec, we search the reasoning step $m$ in $\{1,2,4,6,8,10\}$. 
The coefficient of verifier prediction loss $\beta$ and the strength of monotonicity regularization $\gamma$ are selected from $\{0.3, 0.5, 0.7, 1\}$ and $\{0.1, 0.3, 0.5, 0.7, 1, 1\}$, respectively. 
We implement VRec on a pre-trained LatentR$^3$ for all experiments. 
All experiments are conducted on eight NVIDIA A100-SXM4-80G GPUs. 
More implementation details are provided in anonymous link\footnote{\url{https://anonymous.4open.science/r/VRec/}.}.

\subsection{Overall Performance (RQ1)}\label{sec:overall_performance}
The performance of VRec and baselines are presented in Table~\ref{tab:overall_performance}, from which we can observe that:
\begin{itemize}[leftmargin=*]
    
    \item Among LLM-based generative models, those that utilize human vocabulary (D3 and LatentR3) outperform codebook- (TIGER and LETTER) or embedding-based approaches (SETRec). This is reasonable, as codebook-based methods employ external tokens that are not seen in the LLM pre-training. As such, they typically require substantial data to adapt effectively to the items indexed by the external tokens when scaled to LLMs with billion-scale parameters. Similar observations are also seen in~\citep{bao2024decoding,lin2025order}.
    
    \item Across baselines, LatentR$^3$ consistently surpasses all other methods, demonstrating the effectiveness of the reason-then-recommend paradigm. 
    By introducing an additional attention module dedicated to reasoning and employing reinforcement learning, LatentR$^3$ is encouraged to generate useful reasoning embeddings that facilitates accurate recommendations. It is worth noting that the best performance of LatentR$^3$ occurs with only 1 reasoning step, indicating the lack of scalability on reasoning steps, (\cf analysis in Section~\ref{sec:preliminaries}). This observation is also aligned with the findings reported in~\citep{zhang2025reinforced}.

    \item 
    VRec consistently achieves promising performance across all datasets. 
    In particular, VRec-1Step surpasses baselines in most cases. 
    This is expected due to the incorporation of the verification step, where the verifier can externally evaluate and adjust potentially inaccurate reasoning steps.
    Notably, when extended to multiple steps, VRec attains the best overall performance, demonstrating the scalability and robustness of the verifiable reasoning paradigm. Further analyses of reasoning scalability and a case study on homogeneity alleviation are presented in Section~\ref{sec:reasoning_step_scalability} and Appendix~\ref{app:additiona_results}., respectively.  
    
\end{itemize}

\subsection{In-depth Analysis}

\begin{figure}[t]
% \vspace{-0.2cm}
\setlength{\abovecaptionskip}{0.02cm}
\setlength{\belowcaptionskip}{-0.3cm}
\centering
\includegraphics[width=0.5\linewidth]{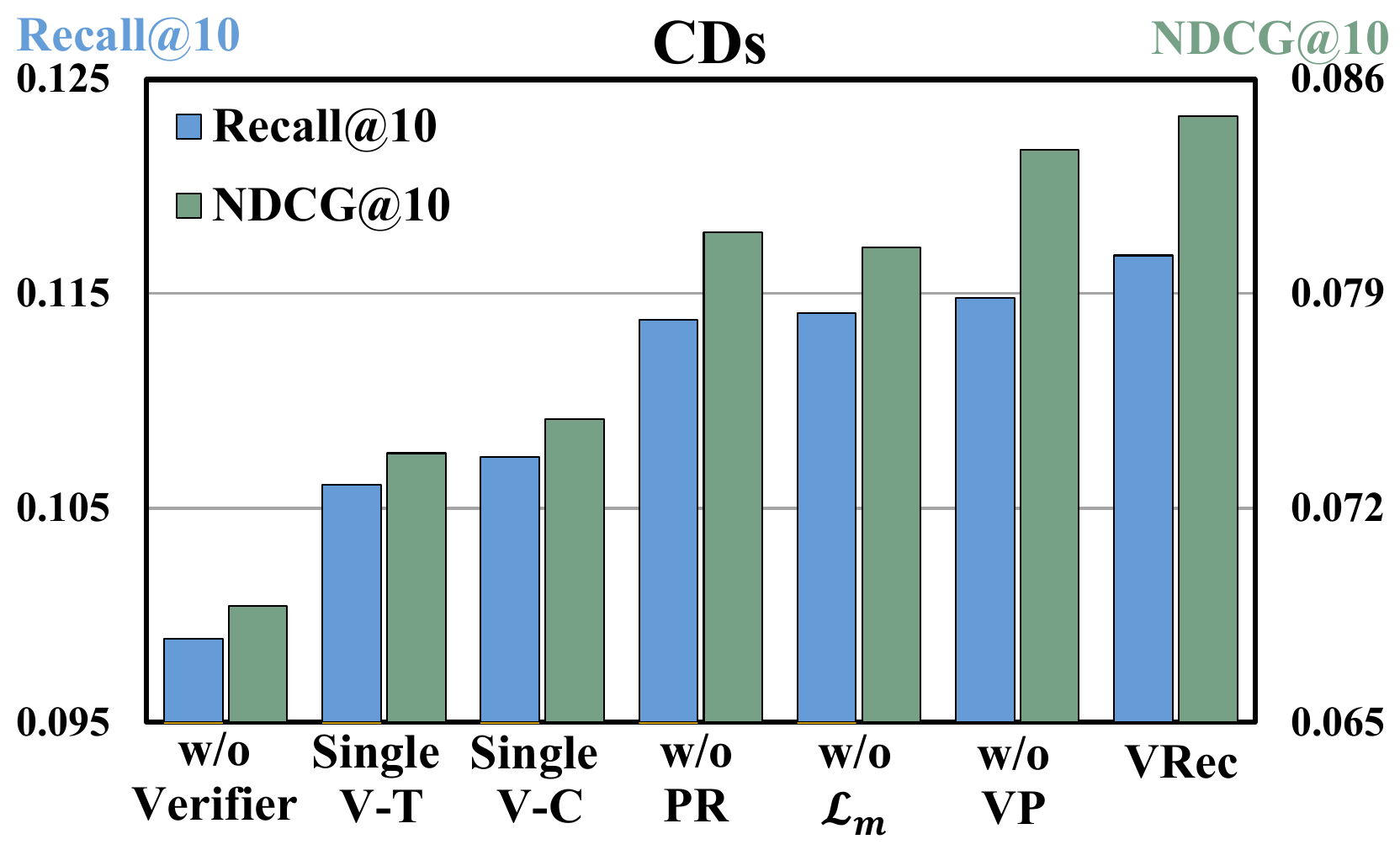}
\caption{Ablation study of VRec on CDs dataset.}
\label{fig:ablation}
\end{figure}

\subsubsection{\textbf{Ablation Study (RQ2)}}
To study the effectiveness of the verification step in VRec, we remove the verification step denoted as ``w/o Verifier''. Besides, to validate the effectiveness of diverse verifiers, we instantiate VRec with only a single verifier from the category and title dimension, denoted as ``Single V-C'' and ``Single V-T'', respectively. 
We also discard the personalized router (``w/o VP) and the monotonicity regularization (``w/o $\mathcal{L}_{m}$''), and remove the verifier pre-training stage (``w/o VP), to evaluate the influence of these components, separately. 

The results on CDs dataset\footnote{We omit results for other datasets with similar observations to save space.} are presented in Figure~\ref{fig:ablation}, where we can find that 
1) removing the verifier that disables the verification step (\ie the LatentR$^3$) leads to a significant performance decline. 
This validates the necessity of incorporating verifier and demonstrates the superiority of the reason-verify-recommend paradigm. 
In addition, 2) employing a single verifier would limit the effectiveness of the verification step, which emphasizes the importance of multi-dimensional verifier to achieve more accurate verification. 
Meanwhile, the inferior performance of VRec without personalized router further indicates the importance of individual-specific verification. 
Furthermore, 
3) removing monotonicity regularization can hurt the performance. This is expected since it can penalize the reasoning that leads to weak preference alignment. 
Lastly,
4) directly training the verifier and LLM recommender jointly leads to inferior recommendation results, which is probably due to the potential conflicts between the task of verifier prediction and the next item prediction if the verifier is trained from scratch.

\subsubsection{\textbf{Reasoning Step Scalability (RQ3)}}\label{sec:reasoning_step_scalability}

\begin{figure}[t]
% \vspace{-0.2cm}
\setlength{\abovecaptionskip}{0.02cm}
\setlength{\belowcaptionskip}{-0.3cm}
\centering
\includegraphics[width=0.5\linewidth]{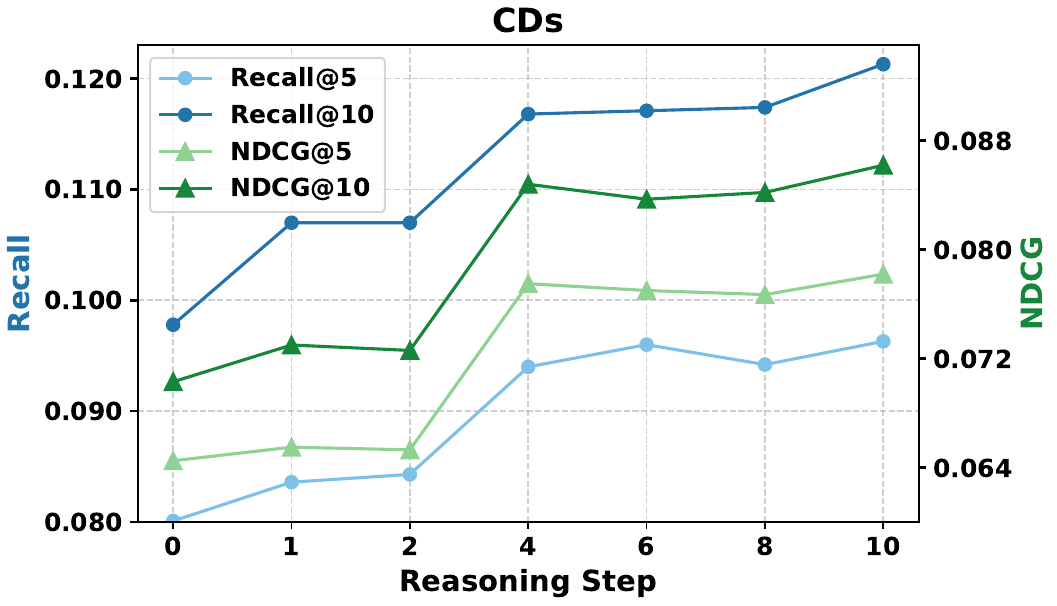}
\caption{Scalability of reasoning steps of VRec.}
\label{fig:reasoning_step_scalability}
\end{figure}

To investigate the potential of the verifiable reasoning paradigm, we vary the number of full reasoning steps $m$ from 0 to 10 and report the results in Figure~\ref{fig:reasoning_step_scalability}. From the results, we can observe that 
1) the recommendation performance gradually improves as the reasoning step increases. This highlights the importance of the verification for the intermediate reasoning, which potentially corrects the reasoning process that misaligns with the user preference. 
Despite the promising test-time scalability of VRec, 2) the performance gains tend to plateau (from {6 to 10 reasoning steps}). This may be because the monotonicity regularization encourages increasing prediction certainty, making it progressively harder for the verifier to achieve more accurate preference alignment as the reasoning steps deepen, thereby limiting significant improvements in later stages. 

\begin{table}[t]
\centering
\setlength{\abovecaptionskip}{0.05cm}
\setlength{\belowcaptionskip}{0.2cm}
\caption{Performance comparison of VRec of different verifier dimensions on CDs dataset.}
\setlength{\tabcolsep}{3mm}{
\resizebox{0.6\textwidth}{!}{
\begin{tabular}{lcccc}
\toprule
 & \multicolumn{1}{l}{\textbf{Recall@5}} & \multicolumn{1}{l}{\textbf{Recall@10}} & \multicolumn{1}{l}{\textbf{NDCG@5}} & \multicolumn{1}{l}{\textbf{NDCG@10}} \\ \midrule
\textbf{Category} & 0.0854 & 0.1074 & 0.0678 & 0.0749 \\
\textbf{Category + Title} & 0.0940 & 0.1168 & 0.0775 & 0.0848 \\
\cellcolor[HTML]{ECF4FF}\textbf{Category + Title + CF} & \cellcolor[HTML]{ECF4FF}0.0973 & \cellcolor[HTML]{ECF4FF}0.1223 & \cellcolor[HTML]{ECF4FF}0.0796 & \cellcolor[HTML]{ECF4FF}0.0877 \\ \bottomrule
\end{tabular}
}}
\label{tab:verifier_diversity}
\end{table}

\begin{figure}[t]
\setlength{\abovecaptionskip}{0.02cm}
\setlength{\belowcaptionskip}{0.3cm}
\centering
\includegraphics[width=0.65\textwidth]{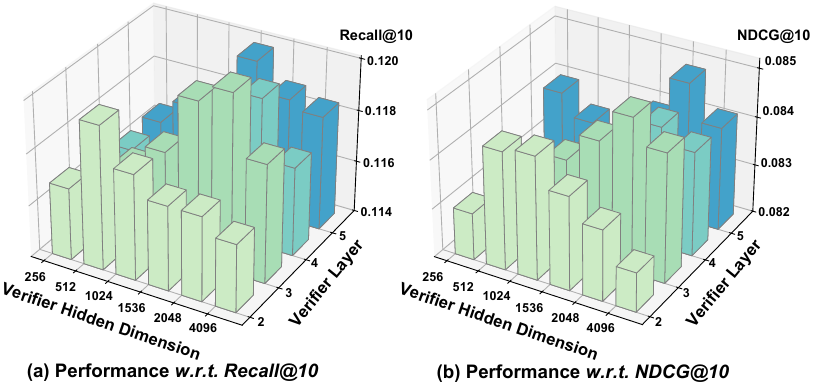}
\caption{Performance comparison of VRec with different verifier model sizes on CDs dataset.}
\label{fig:verifier_size_scalability}
\end{figure}
\subsubsection{\textbf{Verifier  Scalability (RQ3)}}\label{sec:verifier_size_scalability}
% number of verifiers
% verifier model size
In VRec, the key component is the mixture of verifiers, where the model capacity of verifier might also be an important factor for effective verifiable reasoning.  
We investigate the scalability of verifiers from two perspectives:
\begin{itemize}[leftmargin=*]
    \item \textit{Dimensions of verifiers}. 
    % 我们逐步增加verifier的个数，对应的维度分别是category, title, 以及collaborative signals，探究增加维度的影响. 
    % From the results in Table XX, 我们可以看到当逐步扩展verifier的评判维度的时候，performance会持续提升。Notably, semantics的提升相较于cf会提升的更大，这个可能是因为LLM的推理space主要集中在语义部分，而并不擅长CF的捕捉（类似的现象也在xxx中被观测到）
    We gradually increase the number of verifiers, corresponding to the category, title, and collaborative signals, to examine the effect of expanding verification dimensions. From the results in Table~\ref{tab:verifier_diversity}, performance consistently improves as the evaluation dimension broadens. Notably, the improvement from semantic dimensions is more substantial than that from collaborative features, likely because the reasoning space of LLMs is primarily semantic, making them less effective at capturing collaborative patterns. Similar observation has also been reported in prior work~\citep{zhang2025collm}. 
    \item \textit{Model size of verifiers}. 
    % We instantiate the verifier with an MLP and enlarge the hidden dimensions and the hidden layers to explore the scalability of verifier size. 
    We instantiate the verifier with an MLP to explore the verifier's scalability. 
    The results in Figure~\ref{fig:verifier_size_scalability} shows that 
    % modestly把verifier变宽或者变深，能够进一步提高verification的质量，从而让recommendation更准确。但是不断变宽或扩深并不需要，这是因为verifier在做的是high-level的prediction task，类别通常在20左右，不是复杂的reasoning task，这种粗粒度的判别式任务，不需要太复杂，否则反而会overfitting。
    modestly increasing the width or depth of the verifier enhances verification quality, leading to more accurate recommendations, \eg {3 layer with 2048 hidden dimension}. 
    However, continuously enlarging the verifier is unnecessary, as it performs a group-level prediction task with a limited number of classes (around 20). 
    % rather than complex reasoning. 
    Excessive model capacity may therefore introduce overfitting rather than further improvement. 
\end{itemize}

% \subsubsection{\textbf{Generalization Analysis}}
% % gemma 

\begin{table}[t]
\centering
\setlength{\abovecaptionskip}{0.05cm}
\setlength{\belowcaptionskip}{0.2cm}
\caption{Comparison of average inference time costs (per sample) between LLM reasoning with (``w/'') and without (``w/o'') verifier on CDs dataset. }
\setlength{\tabcolsep}{4mm}{
\resizebox{0.7\textwidth}{!}{
\begin{tabular}{lcccccc}
\toprule
\textbf{\# Reasoning Step} & \textbf{1} & \textbf{2} & \textbf{4} & \textbf{6} & \textbf{8} & \textbf{10} \\ \midrule
\textbf{w/o Verifier} & 0.754s & 0.825s & 0.896s & 0.946s & 1.064s & 1.142s \\
\textbf{w/ Verifier} & 0.761s & 0.831s & 0.908s & 0.950s & 1.065s & 1.142s \\
\cellcolor{gray!18}\textbf{\% Time Overhead} & \cellcolor{gray!18}0.97\% & \cellcolor{gray!18}0.71\% & \cellcolor{gray!18}1.33\% & \cellcolor{gray!18}0.46\% & \cellcolor{gray!18}0.06\% & \cellcolor{gray!18}0.03\% \\ \bottomrule
\end{tabular}
}}
\label{tab:efficiency}
\end{table}

\subsubsection{\textbf{Verifier Efficiency Analysis (RQ4)}}\label{sec:efficiency_analysis}
% 0 step, with/ wo verifier 
% To empirically analyze the computational overhead of the verification step to facilitate real-world applications, we test the VRec with and without verification across different reasoning steps. 
% As shown in Figure~\ref{}, 
% incorporating the verification step introduces additional computational overhead as expected. 
% Nonetheless, the computational overhead is neglectable (XXX\% additional time costs on average) as the main time costs are from the LLM itself. 
% As such, we may adjust the reasoning steps to strike the trade-off between the recommendation performance and the time costs from the reasoning steps. 
To empirically assess the computational overhead introduced by the verification step and its practicality in real-world applications, we evaluate VRec with and without verification across different reasoning steps. As shown in Table~\ref{tab:efficiency}, incorporating verification introduces a modest increase in computational cost, as expected. However, the overhead remains negligible (an average of 0.59\% additional time costs), since the majority of the computation is dominated by the LLM backbone. 
As such, the reasoning steps can be flexibly adjusted to balance recommendation performance and inference efficiency, while still benefiting from the performance gains brought by verification.

% \begin{figure}[t]
% \vspace{-0.2cm}
% \setlength{\abovecaptionskip}{-0.15cm}
% \setlength{\belowcaptionskip}{-0.15cm}
%   \centering 
%   \hspace{-0.105in}
%   \subfigure{
%   \includegraphics[height=1.3in]{figures/hp_beta.pdf}} 
%   % \hspace{-0.105in}
%   \subfigure{    
%   \includegraphics[height=1.3in]{figures/hp_gamma.pdf}} 
% \caption{Performance of VRec with different strength of verifier loss $\beta$ and monotonicity regularization $\gamma$.}
% % \CLEAN\HQ{Shall we use different colors for NDCG and Recall (like Fig 6)? nit, $\beta$ and $\gamma$ in the caption seem to be in bold font. }
%   \label{fig:hp_analysis}
%   % \vspace{-0.3cm}
% \end{figure}

% \begin{figure}[t]
% % \vspace{-0.2cm}
% \setlength{\abovecaptionskip}{0cm}
% \setlength{\belowcaptionskip}{0cm}
% \centering
% \begin{subfigure}[b]{0.48\linewidth}
%   \centering
%   \includegraphics[height=1.4in]{figures/hp_beta.pdf}
% \end{subfigure}\hspace{-0.105in}
% \begin{subfigure}[b]{0.48\linewidth}
%   \centering
%   \includegraphics[height=1.4in]{figures/hp_gamma.pdf}
% \end{subfigure}
% \% \caption{Performance of VRec with different strength of verifier loss $\beta$ and monotonicity regularization $\gamma$.}
%   \label{fig:hp_analysis}
% \end{figure}

\begin{figure}[t]
\setlength{\abovecaptionskip}{0cm}
\setlength{\belowcaptionskip}{0cm}
\centering
\begin{subfigure}[b]{0.3\linewidth}
  \centering
  \includegraphics[width=\linewidth]{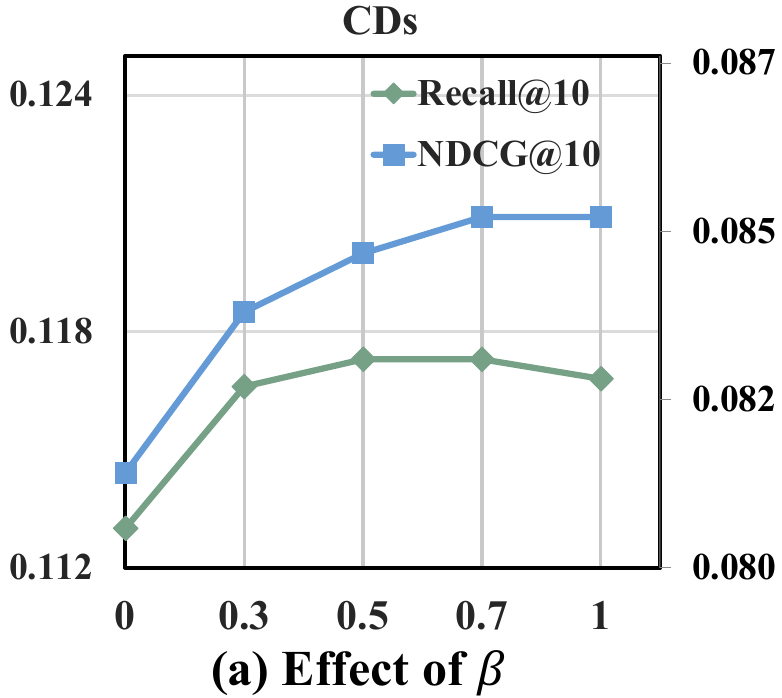}
\end{subfigure}
\hspace{0.01\linewidth}
\begin{subfigure}[b]{0.3\linewidth}
  \centering
  \includegraphics[width=\linewidth]{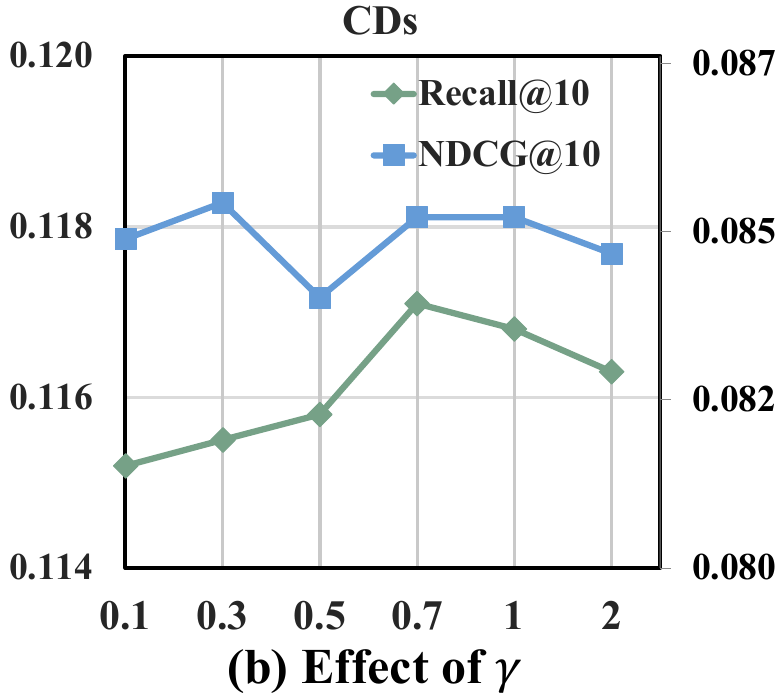}
\end{subfigure}
\caption{Performance of VRec with different strength of verifier loss $\beta$ and monotonicity regularization $\gamma$.}
\label{fig:hp_analysis}
\end{figure}

\subsubsection{\textbf{Hyper-parameter Sensitivity (RQ4)}} 

% alpha - negative sample strength for verifier training
% beta - v_weight
% gamma - mono
% In VRec, we introduce two key hyper-parameters (\ie strength of verifier prediction task $\beta$ and strength of monotonicity regularization). 
We further analyze hyper-parameters of VRec to facilitate future applications. 
% \noindent$\bullet\quad$\textbf{Effect of ${\beta}$}. 
1) \textbf{Effect of verifier loss strength ${\beta}$}. 
From Figure~\ref{fig:hp_analysis}(a), we can observe that the recommendation accuracy first climbs up and then falls as the $\beta$ varies from 0 to 1. 
The verifier needs to maintain sufficient prediction capability to provide effective guidance, but an overly strong verifier may conflict with the recommendation objective. 
% thereby weakening the overall recommendation accuracy. 
2) \textbf{Effect of monotonicity regularization strength ${\gamma}$}. 
% 持续提高，说明了有效性。然后这个monotonicity regularization越提高，它的作用越饱和但不会往下掉，这个是合理的。因为可能当提高到一定程度的时候，它的penalty就忽略了，反而就没有起到更多作用（因为它只要不超过0就行）
As shown in Figure~\ref{fig:hp_analysis}(b), strengthening the monotonicity regularization consistently improves performance, demonstrating its effectiveness in promoting stable reasoning refinement. When $\gamma$ becomes sufficiently large, the improvement gradually saturates without performance degradation. 
This is reasonable since the regularization term only penalizes non-monotonic entropy changes. 
% once the reasoning process becomes stable (\ie the penalty approaches zero), further increasing $\gamma$ yields diminishing influence on optimization. 
% Once the reasoning process becomes stable, further increasing $\gamma$ yields diminishing influence on optimization. 

% \subsubsection{\textbf{Case Study}}

\subsubsection{\textbf{Discussion on Interpretability.}} 
Despite that reasoning in the latent space is very effective and efficient, it is important but challenging to explain the reasoning and verification process. 
Fortunately, our proposed VRec,  has the potential of interpreting the reasoning process with the mixture of verifiers. 
It can leverage the verifiers' group-level predictions to interpret the reasoning process (from ``jazz music'' to ``piano jazz''). 
However, rigorously evaluating the consistency between such explanations and real-world user decision-making is non-trivial, and we leave this investigation for future work. 
% Detailed cases are provided in our anonymous link.

\section{Related Work}\label{sec:related_work}

\subsection{Reasoning for Recommendation}
Recent advances highlight the promise of integrating reasoning into recommendation models~\citep{liu2025recoworld,dai2025onepiece}, where the model performs multiple intermediate reasoning steps before making a final recommendation~\citep{wang2024enhancingrecommendersystemslarge,zhao2025reasontorecommendusinginteractionofthoughtreasoning,xie2025recllmr1twostagetrainingparadigm,gu2025r4ecreasoningreflectionrefinement,xing2025reg4recreasoningenhancedgenerativemodel,huang2025agenticrecommendersystemsera}. 
In LLM-based recommendation, existing efforts can be broadly categorized into explicit and implicit reasoning. 
Explicit approaches typically rely on supervised chain-of-thought (CoT) reasoning and are mostly explored in click-through-rate or rating prediction rather than next-item generation~\citep{xi2024towards,STARec2025,fang2025reason4rec}. However, constructing CoT data is costly, and explicit reasoning often suffers from inefficiency, which poses significant challenges for large-scale LLM deployment. In contrast, implicit (latent) reasoning eliminates the need for annotated CoT data and offers better efficiency~\citep{tang2025thinkrecommendunleashinglatent,lin2026bringing}. Nevertheless, prior studies inevitably encounter unverified reasoning, which leads to homogeneous shortcut reasoning and error accumulation across steps, thereby limiting the effectiveness of reasoning-based recommendations. 
To bridge this gap, our work introduces a novel verifiable reasoning paradigm that provides intermediate feedback at each reasoning step, auditing the reasoning towards a more faithful user preference understanding, unlocking the potential of reasoning.

\subsection{Verifiable LLM Reasoning}  
Verifiable LLM reasoning involves structuring the model's reasoning process so that each intermediate step can be checked for correctness.
Existing studies can be broadly categorized into two groups. 
One line of work explores external-verifier, which explicitly evaluates intermediate reasoning steps using dedicated modules or reward models~\citep{wang2024mathshepherdverifyreinforcellms,zhang2025generativeverifiersrewardmodeling,zhang2025lessonsdevelopingprocessreward,zheng2025processbenchidentifyingprocesserrors}. 
Frameworks such as Self-Refine~\citep{madaan2023selfrefineiterativerefinementselffeedback}, CRITIC~\citep{wang2023drdtdynamicreflectiondivergent,zheng2025criticcotboostingreasoningabilities}, and Tree-of-Thoughts~\citep{yao2023treethoughtsdeliberateproblem} iteratively verify and refine reasoning trajectories through learned critics or search-based evaluators. 
In contrast, 2) internal-signal approaches derive reliability cues from the model's own outputs. Self-Consistency~\citep{wang2023selfconsistencyimproveschainthought,li2023makinglargelanguagemodels,Taubenfeld_2025} aggregates multiple reasoning samples to identify consistent outcomes, while confidence- and entropy-based measures~\citep{wang20258020rulehighentropyminority} quantify uncertainty and calibration. 
However, these methods rely on explicit reasoning traces (\eg textual CoT), making them inefficient and hard to generalize to latent reasoning scenarios in next item recommendation. 
In this work, we propose a general verifiable reasoning framework specifically for LLM-based recommendation that operates on latent reasoning embeddings, achieving step-by-step reasoning verification. 

\section{Conclusion}

In this work, we pointed out the crucial issue of existing reasoning-enhanced LLM-based recommendation, \ie unverified reasoning that can lead to reasoning degradation. 
We then proposed a novel reason-verify-recommend paradigm that introduces an additional verification step to provide reliable feedback guidance for the reasoning process. 
We then summarized two principles for the verifier design, \ie 1) reliable evaluation and guidance feedback, and 2) multi-dimensional verifications. 
Meeting the two principles, we proposed VRec, which employs a mixture of verifiers with a personalized router to ensure multi-dimensional verification. 
To achieve reliable verification, we formulated a group-level preference prediction task for verifiers and leveraged internal model signals (\ie entropy and weights) for verification. 
Empirical results demonstrated the effectiveness, efficiency, and scalability of VRec. 

This work takes the initial attempt to explore verifications for LLM reasoning in next item recommendation, leaving several promising directions. 1) Extending our framework on explicit reasoning for other recommendation tasks, such as CTR or rating prediction, is worth exploring. 2) While efficient, it is important to improve the interpretability and controllability of the verifiable reasoning in latent space,
Third, the design of verifiers presents rich opportunities for exploration, such as developing more expressive architectures and better alignment objectives to strengthen preference modeling and human-aligned verification. 
\section{Appendix}\label{sec:appendix}

% group level inference
% case study

\subsection{Detailed Descriptions}\label{app:detailed_descriptions} 
\noindent$\bullet\quad$\textbf{Detailed analysis of homogeneous reasoning issue.}
Based on the optimization objective, \ie
\begin{equation}
\mathop{\arg\min}_{\theta\in\Theta} \;
\mathbb{E}_{X \sim \mathcal{D}} \;
\mathbb{E}_{R\sim \mathcal{M}_\theta(\cdot\mid X)} \;
\mathcal{L}_\theta(\hat{y}\mid X, R),
\end{equation}
we can observe that the supervision signal is only provided on the final recommendation $\hat{y}$,
while the intermediate reasoning process $R$ is not directly optimized.
As a result, the reasoning model $\mathcal{M}_\theta(R \mid X)$ may converge to a \textit{degenerate distribution}
that collapses into a single deterministic reasoning pattern:
\begin{equation}
\mathcal{M}_\theta(R \mid X) \to \delta_{R^*(X)}(R),
\end{equation}
where $\delta_{R^*(X)}(R)$ denotes a Dirac delta distribution centered at $R^*(X)$,
and $R^*(X) = \arg\max_R \mathcal{M}_\theta(R \mid X)$ represents the most likely (or optimal)
reasoning sequence given input $X$.
In other words, the model tends to generate homogeneous or trivial reasoning steps, lacking diversity across different inputs.

\vspace{2pt}
\noindent$\bullet\quad$\textbf{Training and Procedure Procedure of VRec.} 
We present the overall procedure of VRec instantiation, including two-stage training and inference, in Algorithm~\ref{algo:VRec}.

\noindent$\bullet\quad$\textbf{Group-level User Preference Labeling}. 
To construct the supervision signal for the preference prediction task, we assign each item a group-level label $P$ that reflects coarse-grained user preferences. Specifically, we leverage three types of information in this work: 
1) Category. If the dataset provides item metadata with predefined categories (\ie ``jazz'' or ``hip-hop''), we directly use these as group labels; 
2) Title semantics. We extract semantic embeddings from item titles and perform $k$-means clustering to group items with similar textual content; and 
3) Collaborative information. Based on a pre-trained collaborative filtering (CF) model such as SASRec~\cite{kang2018self}, we cluster item CF embeddings using $k$-means to capture collaborative similarity. The resulting group assignments serve as the preference labels $P$ used for training the verifier in the preference prediction task.

\begin{algorithm}[h]
\small
\caption{Training and Inference Procedure of VRec}
\label{algo:VRec}
\begin{algorithmic}[1]
\Require Pre-trained reasoning LLM recommender parameters $\theta$, initialized verifier parameters $\phi$, training dataset $\mathcal{D}$, total reasoning step $m$, coefficient of verifier loss $\beta$, and coeeficient of monotonicity regularization $\gamma$.

\vspace{3pt}
\Statex \textbf{// Stage 1: Verifier Pre-training}
\ForAll{$(X, y) \in \mathcal{D}$}
    \State Generate reasoning $R$ and recommendation $\hat{y}$ using $\theta$;
    \algorithmiccomment{Eq.(\ref{eqn:reaoning_step})-(\ref{eqn:recommend_step})}
    \State $\mathcal{D}_v \leftarrow \varnothing$
    \If{$\hat{y}=y$}
        \State  $\mathcal{D}_v \leftarrow \mathcal{D}_v \cup
 (R, \mathcal{P})$,  $\mathcal{P}=\{P_1,\dots,P_n\}$; // Positive samples
    \Else
        \State  $\mathcal{D}_v \leftarrow \mathcal{D}_v \cup
 (R, \varnothing)$  // Negative samples
    \EndIf
\EndFor
\State Optimize verifier $\phi$ using $\mathcal{L}_v$ on $\mathcal{D}_v$; 

\vspace{3pt}
\Statex \textbf{// Stage 2: Verifiable Reasoning Fine-tuning}
\ForAll{$(X, y)\in\mathcal{D}$}
    \For{$t=1$ to $m$}
        \State Obtain $\bm{r}_t$ from LLM recommender; 
        \algorithmiccomment{Eq.(\ref{eqn:reaoning_step})}
        \State Obtain $\{f_i, \bm{g}_i\}_{i=1}^{n}$ via Eq.(\ref{Eq:f}); 
        \algorithmiccomment{{Multi-dimensional evaluation}}
        \State 
        Obtain $\bm{r}^{*}_t$ via {Eq.(\ref{Eq:adjustment})}
        \algorithmiccomment{{Confidence-based adjustment}}
        \State  Update $R\leftarrow(R, \bm{r}^{*}_t)$;
        \State Compute $\mathcal{L}_m = \max(0, f_i^{t}-f_i^{t-1})$ via Eq.(\ref{eqn:monotonicity_reg}) ;
        % \algorithmiccomment{Monotonicity regularization}
    \EndFor
    \State Compute recommendation loss $\mathcal{L}_r$; 
    \algorithmiccomment{Eq.(\ref{Eq:llm_loss})}
    \State Update $\theta$ and $\phi$ jointly with:
        $\mathcal{L} = \mathcal{L}_r + \beta\mathcal{L}_v + \gamma\mathcal{L}_m.$
    \algorithmiccomment{Eq.(\ref{Eq:stage3_loss})}
\EndFor

\vspace{3pt}
\Statex \textbf{// Inference}
\For{$t=1$ to $m$}
    \State Perform reasoning step to obtain $\bm{r}_t$;
    \State  $\bm{r^{*}_t} \leftarrow \bm{r}_t$ via Eq.(\ref{Eq:adjustment}); 
    \algorithmiccomment{Confidence-based adjustment}
\EndFor
\State Generate final recommendation $\hat{y}$; 
\algorithmiccomment{Eq.(\ref{eqn:recommend_step})}

\Ensure Final trained parameters $\{\theta, \phi\}$ and recommendation $\hat{y}$.
\end{algorithmic}
\end{algorithm}

\textbf{{Class number for each dimension $d_i$}}. 
Regarding the class number of each verification dimension: 
\begin{itemize}
    \item For category aspect, we adopt the number of the most coarse-grained item category, which is around 20 in the dataset. Specifically, we leverage the category dimension for verification in CDs (27 categories) and Instruments (19 categories) datasets. 
    \item For both title and collaborative dimensions, we empirically select a predefined number for each dataset from $\{15,20,25\}$. The range is aligned with the category dimension, which aims to balance the difficulty of the prediction task across different dimensions. 
\end{itemize}

% dataset statistics
\begin{table}[h]
\centering
\setlength{\abovecaptionskip}{0.05cm}
\setlength{\belowcaptionskip}{0.2cm}
\caption{Datasets statistics. ``Int.'' denotes ``interactions''. }
\setlength{\tabcolsep}{2mm}{
\resizebox{0.48\textwidth}{!}{
\begin{tabular}{lcccc}
\hline
 & \multicolumn{1}{l}{\textbf{\# Train Int.}} & \multicolumn{1}{l}{\textbf{\# Valid Int.}} & \multicolumn{1}{l}{\textbf{\# Test Int.}} & \multicolumn{1}{l}{\textbf{\# Item}} \\ \hline
\textbf{CDs} & 49,251 & 6,156 & 6,158 & 5,841 \\
\textbf{MicroLens} & 60,101 & 9,036 & 9,121 & 5,326 \\
\textbf{Goodreads} & 256,868 & 33,148 & 33,040 & 7,413 \\
\textbf{Instruments} & 66,500 & 8,312 & 8,313 & 5,030 \\ \hline
\end{tabular}
}}
\label{tab:dataset_statistics}
\end{table}

\noindent$\bullet\quad$\textbf{Detailed implementations and settings}. Additional implementation details and experimental settings for the verifier in VRec are as follows. 

\textbf{Verifier number settings.}
To ensure consistent verifier configurations across datasets, we employ two verifiers for overall performance comparison (Section~\ref{sec:overall_performance}. 
Notably, the \textit{Microlens} and \textit{Goodreads} datasets do not contain structured item category information, which limits the number of available group-level preference dimensions. 
To further investigate whether expanding the verifier dimensionality can enhance performance, we conduct an additional analysis in Section~\ref{sec:verifier_size_scalability}. 
The results demonstrate that incorporating more diverse verifier dimensions enables more effective reasoning and leads to performance gains constantly. 

\textbf{Verifier model size settings.} For in-depth analysis of verifier scalability (Section~\ref{sec:verifier_size_scalability}), we use an MLP to implement the verifiers, with hidden dimensions and hidden layers selected in the range of $\{256,512,1024,1536,2048,4096\}$ and $\{2,3,4,5\}$, respectively. For all other experiments, we simply implement the verifier with a linear matrix $W_v\in\mathbb{R}^{d_\mathcal{M}\times d_i}$. 

\textbf{Reasoning step settings.} 
For the in-depth analysis experiments, we observe that VRec’s performance consistently improves as the number of reasoning steps increases from 1 to 10, before gradually reaching saturation. Since increasing the number of reasoning steps inevitably raises test-time computational cost, we adopt a balanced configuration that achieves strong performance while maintaining computational efficiency. Specifically, we set the number of reasoning steps to 4 on the \textit{Amazon CDs} dataset and 6 on the \textit{Microlens} dataset, which provide a favorable trade-off between reasoning effectiveness and inference efficiency.

\subsection{Additional Experimental Results}\label{app:additiona_results}
\begin{figure}[h]
% \vspace{-0.2cm}
\setlength{\abovecaptionskip}{0.02cm}
\setlength{\belowcaptionskip}{-0cm}
\centering
\includegraphics[width=0.5\linewidth]{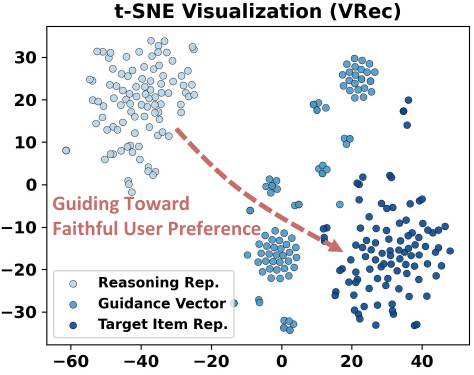}
\caption{Visualization of progressively accurate reasoning of VRec on CDs.}
\label{fig:case_study}
\end{figure}

\noindent$\bullet\quad$\textbf{Case Study}. 
To analyze how VRec alleviates the homogeneous issue, we visualize the reasoning representations, the guidance vectors, and the target item representations via t-SNE visualization as shown in Figure~\ref{fig:case_study}. 
The visualization demonstrates a clear guiding trend from the initial reasoning representations towards the target item representations. 
This validates the effectiveness of VRec in alleviating the reasoning degradation issue, thus leading to more appropriate recommendations. 

\begin{figure}[h]
% \vspace{-0.2cm}
\setlength{\abovecaptionskip}{0.02cm}
\setlength{\belowcaptionskip}{-0cm}
\centering
\includegraphics[width=0.5\linewidth]{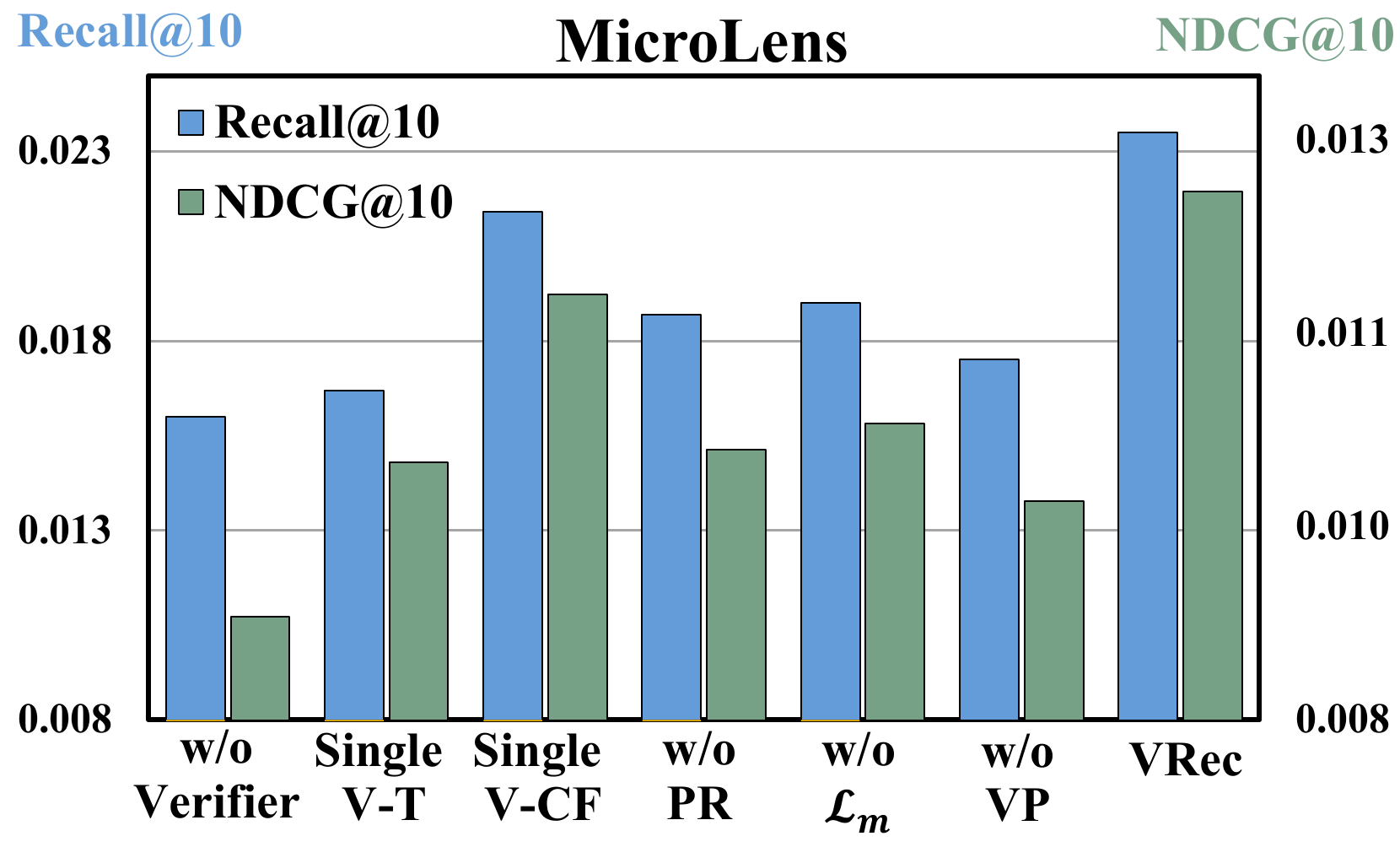}
\caption{Scalability of reasoning steps.}
\label{fig:ablaion-microlens}
\end{figure}

\vspace{2pt}
\noindent$\bullet\quad$\textbf{Ablation study on MicroLens.} The ablation study results are shown in Figure~\ref{fig:ablaion-microlens}, from which we have similar observations with that on CDs, including the most significant influence of verifier (the worst performance of ``w/o Verifier'') and the effectiveness of each component in VRec. 
Nonetheless, we find another interesting observation: verifying the reasoning from the collaborative dimension (``Single V-CF'') is more helpful than verifying from the semantic dimension (``Single V-T''). 
A possible reason is that the textual descriptions in MicroLens are noisy, \eg a very long sentence irrelevant to the main content of the micro-videos. 
In this case, the collaborative information can be useful to represent some important item features, such as popularity.

\begin{figure}[h]
% \vspace{-0.2cm}
\setlength{\abovecaptionskip}{0.02cm}
\setlength{\belowcaptionskip}{-0.3cm}
\centering
\includegraphics[width=0.5\linewidth]{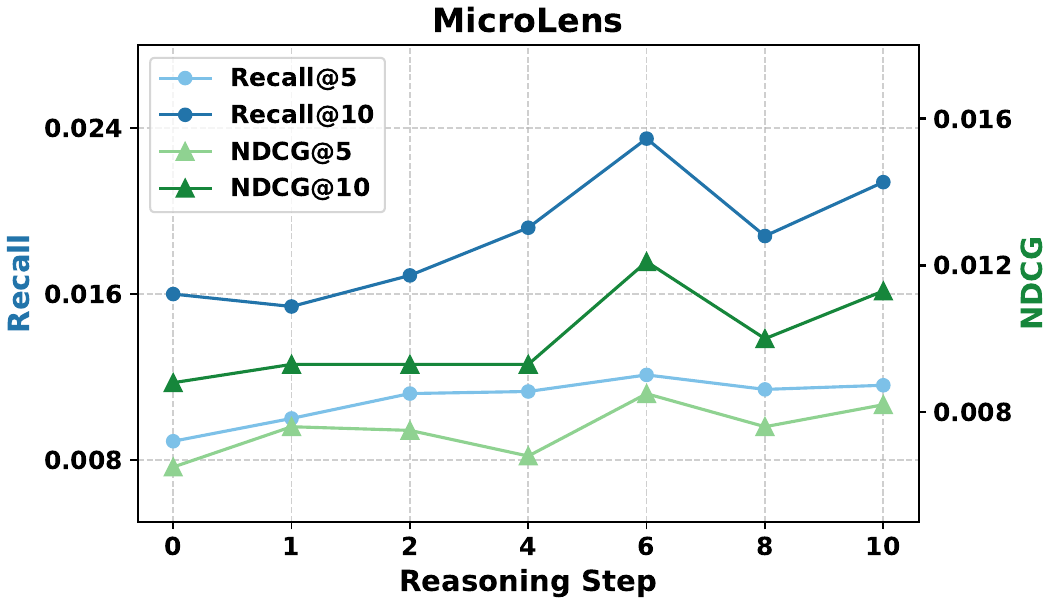}
\caption{Scalability of reasoning steps.}
\label{fig:app_step_scalability_microlens}
\end{figure}

\noindent$\bullet\quad$\textbf{Reasoning Step Scalability on MicroLens dataset.} 
The performance of VRec with different reasoning steps $m$ on MicroLens dataset is presented in Figure~\ref{fig:app_step_scalability_microlens}. From the results, we have similar observations with those on CDs dataset, which demonstrates the superiority of VRec in mitigating the reasoning degradation issue, thus facilitating promising scalability test-time reasoning steps.

\bibliographystyle{assets/plainnat}
\bibliography{custom}

\begin{thebibliography}{63}
\providecommand{\natexlab}[1]{#1}
\providecommand{\url}[1]{\texttt{#1}}
\expandafter\ifx\csname urlstyle\endcsname\relax
  \providecommand{\doi}[1]{doi: #1}\else
  \providecommand{\doi}{doi: \begingroup \urlstyle{rm}\Url}\fi

\bibitem[Bao et~al.(2024)Bao, Zhang, Zhang, Huo, Chen, and Feng]{bao2024decoding}
Keqin Bao, Jizhi Zhang, Yang Zhang, Xinyue Huo, Chong Chen, and Fuli Feng.
\newblock Decoding matters: Addressing amplification bias and homogeneity issue in recommendations for large language models.
\newblock In \emph{ACL}. ACL, 2024.

\bibitem[Bao et~al.(2025)Bao, Zhang, Wang, Zhang, Yang, Luo, Chen, Feng, and Tian]{bao2025bi}
Keqin Bao, Jizhi Zhang, Wenjie Wang, Yang Zhang, Zhengyi Yang, Yanchen Luo, Chong Chen, Fuli Feng, and Qi~Tian.
\newblock A bi-step grounding paradigm for large language models in recommendation systems.
\newblock \emph{ACM Transactions on Recommender Systems}, 3\penalty0 (4):\penalty0 1--27, 2025.

\bibitem[Dai et~al.(2025)Dai, Tang, Wu, Wang, Zhu, Chen, Hong, Zhao, Fu, Wu, et~al.]{dai2025onepiece}
Sunhao Dai, Jiakai Tang, Jiahua Wu, Kun Wang, Yuxuan Zhu, Bingjun Chen, Bangyang Hong, Yu~Zhao, Cong Fu, Kangle Wu, et~al.
\newblock Onepiece: Bringing context engineering and reasoning to industrial cascade ranking system.
\newblock \emph{arXiv preprint arXiv:2509.18091}, 2025.

\bibitem[Ding et~al.(2024{\natexlab{a}})Ding, Fan, Guehring, Gupta, Ha, Huan, Liu, Omidvar-Tehrani, Wang, and Zhou]{ding2024reasoning}
Hao Ding, Ziwei Fan, Ingo Guehring, Gaurav Gupta, Wooseok Ha, Jun Huan, Linbo Liu, Behrooz Omidvar-Tehrani, Shiqi Wang, and Hao Zhou.
\newblock Reasoning and planning with large language models in code development.
\newblock In \emph{KDD}, pages 6480--6490, 2024{\natexlab{a}}.

\bibitem[Ding et~al.(2024{\natexlab{b}})Ding, Liu, Fu, Song, Xie, and Zhang]{ding2024break}
Mengru Ding, Hanmeng Liu, Zhizhang Fu, Jian Song, Wenbo Xie, and Yue Zhang.
\newblock Break the chain: Large language models can be shortcut reasoners.
\newblock \emph{arXiv:2406.06580}, 2024{\natexlab{b}}.

\bibitem[Fang et~al.(2025)Fang, Wang, Zhang, Zhu, Wang, Feng, and He]{fang2025reason4rec}
Yi~Fang, Wenjie Wang, Yang Zhang, Fengbin Zhu, Qifan Wang, Fuli Feng, and Xiangnan He.
\newblock Reason4rec: Large language models for recommendation with deliberative user preference alignment.
\newblock \emph{arXiv preprint arXiv:2502.02061}, 2025.

\bibitem[Farquhar et~al.(2024)Farquhar, Kossen, Kuhn, and Gal]{farquhar2024detecting}
Sebastian Farquhar, Jannik Kossen, Lorenz Kuhn, and Yarin Gal.
\newblock Detecting hallucinations in large language models using semantic entropy.
\newblock \emph{Nature}, 630\penalty0 (8017):\penalty0 625--630, 2024.

\bibitem[Fu et~al.(2025)Fu, Wang, Tian, and Zhao]{fu2025deep}
Yichao Fu, Xuewei Wang, Yuandong Tian, and Jiawei Zhao.
\newblock Deep think with confidence.
\newblock \emph{arXiv:2508.15260}, 2025.

\bibitem[Geirhos et~al.(2020)Geirhos, Jacobsen, Michaelis, Zemel, Brendel, Bethge, and Wichmann]{geirhos2020shortcut}
Robert Geirhos, J{\"o}rn-Henrik Jacobsen, Claudio Michaelis, Richard Zemel, Wieland Brendel, Matthias Bethge, and Felix~A Wichmann.
\newblock Shortcut learning in deep neural networks.
\newblock \emph{Nature Machine Intelligence}, 2\penalty0 (11):\penalty0 665--673, 2020.

\bibitem[Gu et~al.(2025)Gu, Zhong, Xia, Yang, Lu, Jiang, and Gai]{gu2025r4ecreasoningreflectionrefinement}
Hao Gu, Rui Zhong, Yu~Xia, Wei Yang, Chi Lu, Peng Jiang, and Kun Gai.
\newblock R4ec: A reasoning, reflection, and refinement framework for recommendation systems, 2025.
\newblock \url{https://arxiv.org/abs/2507.17249}.

\bibitem[Guo et~al.(2025)Guo, Yang, Zhang, Song, Zhang, Xu, Zhu, Ma, Wang, Bi, et~al.]{guo2025deepseek}
Daya Guo, Dejian Yang, Haowei Zhang, Junxiao Song, Ruoyu Zhang, Runxin Xu, Qihao Zhu, Shirong Ma, Peiyi Wang, Xiao Bi, et~al.
\newblock Deepseek-r1: Incentivizing reasoning capability in llms via reinforcement learning.
\newblock \emph{arXiv:2501.12948}, 2025.

\bibitem[Hidasi et~al.(2016)Hidasi, Karatzoglou, Baltrunas, and Tikk]{hidasi2015session}
Bal{\'a}zs Hidasi, Alexandros Karatzoglou, Linas Baltrunas, and Domonkos Tikk.
\newblock Session-based recommendations with recurrent neural networks.
\newblock In \emph{ICLR}, 2016.

\bibitem[Huang et~al.(2025)Huang, Wu, Xia, Yu, Wang, Yu, Zhang, Rossi, Kveton, Zhou, McAuley, and Yao]{huang2025agenticrecommendersystemsera}
Chengkai Huang, Junda Wu, Yu~Xia, Zixu Yu, Ruhan Wang, Tong Yu, Ruiyi Zhang, Ryan~A. Rossi, Branislav Kveton, Dongruo Zhou, Julian McAuley, and Lina Yao.
\newblock Towards agentic recommender systems in the era of multimodal large language models, 2025.
\newblock \url{https://arxiv.org/abs/2503.16734}.

\bibitem[Huang and Yang(2025)]{huang2025winninggoldimo2025}
Yichen Huang and Lin~F. Yang.
\newblock Winning gold at imo 2025 with a model-agnostic verification-and-refinement pipeline, 2025.

\bibitem[Kang and McAuley(2018)]{kang2018self}
Wang-Cheng Kang and Julian McAuley.
\newblock Self-attentive sequential recommendation.
\newblock In \emph{ICDM}, pages 197--206. IEEE, 2018.

\bibitem[Kim et~al.()Kim, Kim, Cho, Kang, Chang, Yeo, and Lee]{kim2408review}
Jieyong Kim, Hyunseo Kim, Hyunjin Cho, SeongKu Kang, Buru Chang, Jinyoung Yeo, and Dongha Lee.
\newblock Review-driven personalized preference reasoning with large language models for recommendation. corr, abs/2408.06276, 2024. doi: 10.48550.
\newblock \emph{arXiv preprint ARXIV.2408.06276}.

\bibitem[Li et~al.(2018)Li, Liu, Chen, and Rudin]{li2018deep}
Oscar Li, Hao Liu, Chaofan Chen, and Cynthia Rudin.
\newblock Deep learning for case-based reasoning through prototypes: A neural network that explains its predictions.
\newblock In \emph{AAAI}, volume~32, 2018.

\bibitem[Li et~al.(2023)Li, Lin, Zhang, Fu, Chen, Lou, and Chen]{li2023makinglargelanguagemodels}
Yifei Li, Zeqi Lin, Shizhuo Zhang, Qiang Fu, Bei Chen, Jian-Guang Lou, and Weizhu Chen.
\newblock Making large language models better reasoners with step-aware verifier, 2023.
\newblock \url{https://arxiv.org/abs/2206.02336}.

\bibitem[Lin et~al.(2024)Lin, Wang, Li, Feng, Ng, and Chua]{lin2024bridging}
Xinyu Lin, Wenjie Wang, Yongqi Li, Fuli Feng, See-Kiong Ng, and Tat-Seng Chua.
\newblock Bridging items and language: A transition paradigm for large language model-based recommendation.
\newblock In \emph{KDD}, pages 1816--1826. ACM, 2024.

\bibitem[Lin et~al.(2025)Lin, Shi, Wang, Feng, Wang, Ng, and Chua]{lin2025order}
Xinyu Lin, Haihan Shi, Wenjie Wang, Fuli Feng, Qifan Wang, See-Kiong Ng, and Tat-Seng Chua.
\newblock Order-agnostic identifier for large language model-based generative recommendation.
\newblock In \emph{SIGIR}, pages 1923--1933, 2025.

\bibitem[Lin et~al.(2026)Lin, Liu, Wang, Hu, Xu, Feng, Wang, and Chua]{lin2026bringing}
Xinyu Lin, Pengyuan Liu, Wenjie Wang, Yicheng Hu, Chen Xu, Fuli Feng, Qifan Wang, and Tat-Seng Chua.
\newblock Bringing reasoning to generative recommendation through the lens of cascaded ranking.
\newblock \emph{arXiv:2602.03692}, 2026.

\bibitem[Liu et~al.(2025)Liu, Lin, Yu, Wu, Wang, Zhang, Zhao, Xia, Zhang, Li, et~al.]{liu2025recoworld}
Fei Liu, Xinyu Lin, Hanchao Yu, Mingyuan Wu, Jianyu Wang, Qiang Zhang, Zhuokai Zhao, Yinglong Xia, Yao Zhang, Weiwei Li, et~al.
\newblock Recoworld: Building simulated environments for agentic recommender systems.
\newblock In \emph{arXiv:2509.10397}, 2025.

\bibitem[Liu et~al.(2024)Liu, Dou, Wang, Peng, and Yue]{liu2024uncertainty}
Hao Liu, Zi-Yi Dou, Yixin Wang, Nanyun Peng, and Yisong Yue.
\newblock Uncertainty calibration for tool-using language agents.
\newblock In \emph{EMNLP}, pages 16781--16805, 2024.

\bibitem[Madaan et~al.(2023)Madaan, Tandon, Gupta, Hallinan, Gao, Wiegreffe, Alon, Dziri, Prabhumoye, Yang, Gupta, Majumder, Hermann, Welleck, Yazdanbakhsh, and Clark]{madaan2023selfrefineiterativerefinementselffeedback}
Aman Madaan, Niket Tandon, Prakhar Gupta, Skyler Hallinan, Luyu Gao, Sarah Wiegreffe, Uri Alon, Nouha Dziri, Shrimai Prabhumoye, Yiming Yang, Shashank Gupta, Bodhisattwa~Prasad Majumder, Katherine Hermann, Sean Welleck, Amir Yazdanbakhsh, and Peter Clark.
\newblock Self-refine: Iterative refinement with self-feedback, 2023.
\newblock \url{https://arxiv.org/abs/2303.17651}.

\bibitem[Morin and Bengio(2005)]{morin2005hierarchical}
Frederic Morin and Yoshua Bengio.
\newblock Hierarchical probabilistic neural network language model.
\newblock In \emph{International workshop on artificial intelligence and statistics}, pages 246--252. PMLR, 2005.

\bibitem[Ni et~al.(2023)Ni, Cheng, Liu, Fu, Li, He, Zhang, and Yuan]{ni2023content}
Yongxin Ni, Yu~Cheng, Xiangyan Liu, Junchen Fu, Youhua Li, Xiangnan He, Yongfeng Zhang, and Fajie Yuan.
\newblock A content-driven micro-video recommendation dataset at scale.
\newblock \emph{arXiv:2309.15379}, 2023.

\bibitem[OpenAI(2025)]{openai2025reasoning}
OpenAI.
\newblock Learning to reason with llms, 2025.
\newblock \url{https://openai.com/index/learning-to-reason-with-llms/}.

\bibitem[Qwen et~al.(2025)Qwen, :, Yang, Yang, Zhang, Hui, Zheng, Yu, Li, Liu, Huang, Wei, Lin, Yang, Tu, Zhang, Yang, Yang, Zhou, Lin, Dang, Lu, Bao, Yang, Yu, Li, Xue, Zhang, Zhu, Men, Lin, Li, Tang, Xia, Ren, Ren, Fan, Su, Zhang, Wan, Liu, Cui, Zhang, and Qiu]{qwen2025qwen25technicalreport}
Qwen, :, An~Yang, Baosong Yang, Beichen Zhang, Binyuan Hui, Bo~Zheng, Bowen Yu, Chengyuan Li, Dayiheng Liu, Fei Huang, Haoran Wei, Huan Lin, Jian Yang, Jianhong Tu, Jianwei Zhang, Jianxin Yang, Jiaxi Yang, Jingren Zhou, Junyang Lin, Kai Dang, Keming Lu, Keqin Bao, Kexin Yang, Le~Yu, Mei Li, Mingfeng Xue, Pei Zhang, Qin Zhu, Rui Men, Runji Lin, Tianhao Li, Tianyi Tang, Tingyu Xia, Xingzhang Ren, Xuancheng Ren, Yang Fan, Yang Su, Yichang Zhang, Yu~Wan, Yuqiong Liu, Zeyu Cui, Zhenru Zhang, and Zihan Qiu.
\newblock Qwen2.5 technical report.
\newblock 2025.

\bibitem[Rajput et~al.(2023)Rajput, Mehta, Singh, Keshavan, Vu, Heldt, Hong, Tay, Tran, Samost, et~al.]{rajput2023recommender}
Shashank Rajput, Nikhil Mehta, Anima Singh, Raghunandan~H Keshavan, Trung Vu, Lukasz Heldt, Lichan Hong, Yi~Tay, Vinh~Q Tran, Jonah Samost, et~al.
\newblock Recommender systems with generative retrieval.
\newblock In \emph{NeurIPS}. Curran Associates, Inc., 2023.

\bibitem[Tang et~al.(2025{\natexlab{a}})Tang, Dai, Shi, Xu, Chen, Chen, Wu, and Jiang]{tang2025think}
Jiakai Tang, Sunhao Dai, Teng Shi, Jun Xu, Xu~Chen, Wen Chen, Jian Wu, and Yuning Jiang.
\newblock Think before recommend: Unleashing the latent reasoning power for sequential recommendation.
\newblock \emph{arXiv preprint arXiv:2503.22675}, 2025{\natexlab{a}}.

\bibitem[Tang et~al.(2025{\natexlab{b}})Tang, Dai, Shi, Xu, Chen, Chen, Wu, and Jiang]{tang2025thinkrecommendunleashinglatent}
Jiakai Tang, Sunhao Dai, Teng Shi, Jun Xu, Xu~Chen, Wen Chen, Jian Wu, and Yuning Jiang.
\newblock Think before recommend: Unleashing the latent reasoning power for sequential recommendation, 2025{\natexlab{b}}.
\newblock \url{https://arxiv.org/abs/2503.22675}.

\bibitem[Tang and Wang(2018)]{tang2018personalized}
Jiaxi Tang and Ke~Wang.
\newblock Personalized top-n sequential recommendation via convolutional sequence embedding.
\newblock In \emph{WSDM}, pages 565--573, 2018.

\bibitem[Taubenfeld et~al.(2025)Taubenfeld, Sheffer, Ofek, Feder, Goldstein, Gekhman, and Yona]{Taubenfeld_2025}
Amir Taubenfeld, Tom Sheffer, Eran Ofek, Amir Feder, Ariel Goldstein, Zorik Gekhman, and Gal Yona.
\newblock Confidence improves self-consistency in llms.
\newblock In \emph{Findings of the Association for Computational Linguistics: ACL 2025}, page 20090–20111. Association for Computational Linguistics, 2025.
\newblock \doi{10.18653/v1/2025.findings-acl.1030}.
\newblock \url{http://dx.doi.org/10.18653/v1/2025.findings-acl.1030}.

\bibitem[Team et~al.(2025)Team, Bai, Bao, Chen, Chen, Chen, Chen, Chen, Chen, Chen, et~al.]{team2025kimi}
Kimi Team, Yifan Bai, Yiping Bao, Guanduo Chen, Jiahao Chen, Ningxin Chen, Ruijue Chen, Yanru Chen, Yuankun Chen, Yutian Chen, et~al.
\newblock Kimi k2: Open agentic intelligence.
\newblock \emph{arXiv:2507.20534}, 2025.

\bibitem[Team et~al.(2024)]{team2024qwen2}
Qwen Team et~al.
\newblock Qwen2 technical report.
\newblock \emph{arXiv preprint arXiv:2407.10671}, 2:\penalty0 3, 2024.

\bibitem[Tsai et~al.(2024)Tsai, Kraft, Jin, Cai, Hosseini, Xu, Zhang, Hong, Chi, and Yi]{tsai2024leveraging}
Alicia~Y Tsai, Adam Kraft, Long Jin, Chenwei Cai, Anahita Hosseini, Taibai Xu, Zemin Zhang, Lichan Hong, Ed~H Chi, and Xinyang Yi.
\newblock Leveraging llm reasoning enhances personalized recommender systems.
\newblock \emph{arXiv preprint arXiv:2408.00802}, 2024.

\bibitem[Wang et~al.(2024{\natexlab{a}})Wang, Li, Shao, Xu, Dai, Li, Chen, Wu, and Sui]{wang2024mathshepherdverifyreinforcellms}
Peiyi Wang, Lei Li, Zhihong Shao, R.~X. Xu, Damai Dai, Yifei Li, Deli Chen, Y.~Wu, and Zhifang Sui.
\newblock Math-shepherd: Verify and reinforce llms step-by-step without human annotations, 2024{\natexlab{a}}.
\newblock \url{https://arxiv.org/abs/2312.08935}.

\bibitem[Wang et~al.(2025)Wang, Yu, Gao, Zheng, Liu, Lu, Dang, Chen, Yang, Zhang, Liu, Yang, Zhao, Yue, Song, Yu, Huang, and Lin]{wang20258020rulehighentropyminority}
Shenzhi Wang, Le~Yu, Chang Gao, Chujie Zheng, Shixuan Liu, Rui Lu, Kai Dang, Xionghui Chen, Jianxin Yang, Zhenru Zhang, Yuqiong Liu, An~Yang, Andrew Zhao, Yang Yue, Shiji Song, Bowen Yu, Gao Huang, and Junyang Lin.
\newblock Beyond the 80/20 rule: High-entropy minority tokens drive effective reinforcement learning for llm reasoning, 2025.
\newblock \url{https://arxiv.org/abs/2506.01939}.

\bibitem[Wang et~al.(2021)Wang, Feng, He, Nie, and Chua]{wang2021denoising}
Wenjie Wang, Fuli Feng, Xiangnan He, Liqiang Nie, and Tat-Seng Chua.
\newblock Denoising implicit feedback for recommendation.
\newblock In \emph{WSDM}, pages 373--381. ACM, 2021.

\bibitem[Wang et~al.(2024{\natexlab{b}})Wang, Bao, Lin, Zhang, Li, Feng, Ng, and Chua]{wang2024learnable}
Wenjie Wang, Honghui Bao, Xinyu Lin, Jizhi Zhang, Yongqi Li, Fuli Feng, See-Kiong Ng, and Tat-Seng Chua.
\newblock Learnable item tokenization for generative recommendation.
\newblock In \emph{CIKM}, pages 2400--2409, 2024{\natexlab{b}}.

\bibitem[Wang et~al.(2023{\natexlab{a}})Wang, Wei, Schuurmans, Le, Chi, Narang, Chowdhery, and Zhou]{wang2023selfconsistencyimproveschainthought}
Xuezhi Wang, Jason Wei, Dale Schuurmans, Quoc Le, Ed~Chi, Sharan Narang, Aakanksha Chowdhery, and Denny Zhou.
\newblock Self-consistency improves chain of thought reasoning in language models, 2023{\natexlab{a}}.
\newblock \url{https://arxiv.org/abs/2203.11171}.

\bibitem[Wang et~al.(2024{\natexlab{c}})Wang, Chu, Ouyang, Wang, Hao, Shen, Gu, Xue, Zhang, Cui, Li, Zhou, and Li]{wang2024enhancingrecommendersystemslarge}
Yan Wang, Zhixuan Chu, Xin Ouyang, Simeng Wang, Hongyan Hao, Yue Shen, Jinjie Gu, Siqiao Xue, James~Y Zhang, Qing Cui, Longfei Li, Jun Zhou, and Sheng Li.
\newblock Enhancing recommender systems with large language model reasoning graphs, 2024{\natexlab{c}}.
\newblock \url{https://arxiv.org/abs/2308.10835}.

\bibitem[Wang et~al.(2023{\natexlab{b}})Wang, Liu, Zhang, Yao, Heinecke, and Yu]{wang2023drdtdynamicreflectiondivergent}
Yu~Wang, Zhiwei Liu, Jianguo Zhang, Weiran Yao, Shelby Heinecke, and Philip~S. Yu.
\newblock Drdt: Dynamic reflection with divergent thinking for llm-based sequential recommendation, 2023{\natexlab{b}}.
\newblock \url{https://arxiv.org/abs/2312.11336}.

\bibitem[Wei et~al.(2025)Wei, Zhang, He, Xia, Pan, and Liu]{wei2025plangenllms}
Hui Wei, Zihao Zhang, Shenghua He, Tian Xia, Shijia Pan, and Fei Liu.
\newblock Plangenllms: A modern survey of llm planning capabilities.
\newblock ACL, 2025.

\bibitem[Wei et~al.(2022)Wei, Wang, Schuurmans, Bosma, Xia, Chi, Le, Zhou, et~al.]{wei2022chain}
Jason Wei, Xuezhi Wang, Dale Schuurmans, Maarten Bosma, Fei Xia, Ed~Chi, Quoc~V Le, Denny Zhou, et~al.
\newblock Chain-of-thought prompting elicits reasoning in large language models.
\newblock volume~35, pages 24824--24837, 2022.

\bibitem[Wu et~al.(2025)Wu, Ren, Zhang, Wang, Ma, Ye, and Zhao]{STARec2025}
Chenghao Wu, Ruiyang Ren, Junjie Zhang, Ruirui Wang, Zhongrui Ma, Qi~Ye, and Wayne~Xin Zhao.
\newblock {STARec}: An efficient agent framework for recommender systems via autonomous deliberate reasoning.
\newblock In \emph{Proceedings of the 34th ACM International Conference on Information and Knowledge Management (CIKM)}, 2025.
\newblock \url{https://arxiv.org/abs/2508.18812}.

\bibitem[Xi et~al.(2024)Xi, Liu, Lin, Cai, Zhu, Zhu, Chen, Tang, Zhang, and Yu]{xi2024towards}
Yunjia Xi, Weiwen Liu, Jianghao Lin, Xiaoling Cai, Hong Zhu, Jieming Zhu, Bo~Chen, Ruiming Tang, Weinan Zhang, and Yong Yu.
\newblock Towards open-world recommendation with knowledge augmentation from large language models.
\newblock In \emph{RecSys}, pages 12--22, 2024.

\bibitem[Xie et~al.(2025)Xie, Ren, Qi, Hu, and Shan]{xie2025recllmr1twostagetrainingparadigm}
Yu~Xie, Xingkai Ren, Ying Qi, Yao Hu, and Lianlei Shan.
\newblock Recllm-r1: A two-stage training paradigm with reinforcement learning and chain-of-thought v1, 2025.
\newblock \url{https://arxiv.org/abs/2506.19235}.

\bibitem[Xing et~al.(2025)Xing, Deng, Mao, Hu, Xu, Zhang, Wang, Wang, Zhang, Zeng, and Zhang]{xing2025reg4recreasoningenhancedgenerativemodel}
Haibo Xing, Hao Deng, Yucheng Mao, Jinxin Hu, Yi~Xu, Hao Zhang, Jiahao Wang, Shizhun Wang, Yu~Zhang, Xiaoyi Zeng, and Jing Zhang.
\newblock Reg4rec: Reasoning-enhanced generative model for large-scale recommendation systems, 2025.
\newblock \url{https://arxiv.org/abs/2508.15308}.

\bibitem[Yao et~al.(2023)Yao, Yu, Zhao, Shafran, Griffiths, Cao, and Narasimhan]{yao2023treethoughtsdeliberateproblem}
Shunyu Yao, Dian Yu, Jeffrey Zhao, Izhak Shafran, Thomas~L. Griffiths, Yuan Cao, and Karthik Narasimhan.
\newblock Tree of thoughts: Deliberate problem solving with large language models, 2023.
\newblock \url{https://arxiv.org/abs/2305.10601}.

\bibitem[You et~al.(2025{\natexlab{a}})You, Li, Lin, Zhang, Wang, Li, and Nie]{you2025text}
Runyang You, Yongqi Li, Xinyu Lin, Xin Zhang, Wenjie Wang, Wenjie Li, and Liqiang Nie.
\newblock {R}$^2${ec}: Towards large recommender models with reasoning.
\newblock In \emph{NeurIPS}, 2025{\natexlab{a}}.

\bibitem[You et~al.(2025{\natexlab{b}})You, Li, Liu, Wang, Nie, and Li]{you2025parallel}
Runyang You, Yongqi Li, Meng Liu, Wenjie Wang, Liqiang Nie, and Wenjie Li.
\newblock Parallel test-time scaling for latent reasoning models.
\newblock \emph{arXiv:2510.07745}, 2025{\natexlab{b}}.

\bibitem[Yuan et~al.(2024)Yuan, Zhao, Zhang, Zheng, and Liu]{yuan2024llms}
Yu~Yuan, Lili Zhao, Kai Zhang, Guangting Zheng, and Qi~Liu.
\newblock Do llms overcome shortcut learning? an evaluation of shortcut challenges in large language models.
\newblock In \emph{EMNLP}, pages 12188--12200, 2024.

\bibitem[Zhang et~al.(2025{\natexlab{a}})Zhang, Hosseini, Bansal, Kazemi, Kumar, and Agarwal]{zhang2025generativeverifiersrewardmodeling}
Lunjun Zhang, Arian Hosseini, Hritik Bansal, Mehran Kazemi, Aviral Kumar, and Rishabh Agarwal.
\newblock Generative verifiers: Reward modeling as next-token prediction, 2025{\natexlab{a}}.
\newblock \url{https://arxiv.org/abs/2408.15240}.

\bibitem[Zhang et~al.(2025{\natexlab{b}})Zhang, Feng, Zhang, Bao, Wang, and He]{zhang2025collm}
Yang Zhang, Fuli Feng, Jizhi Zhang, Keqin Bao, Qifan Wang, and Xiangnan He.
\newblock Collm: Integrating collaborative embeddings into large language models for recommendation.
\newblock \emph{TKDE}, 2025{\natexlab{b}}.

\bibitem[Zhang et~al.(2025{\natexlab{c}})Zhang, Xu, Zhao, Wang, Feng, He, and Chua]{zhang2025reinforced}
Yang Zhang, Wenxin Xu, Xiaoyan Zhao, Wenjie Wang, Fuli Feng, Xiangnan He, and Tat-Seng Chua.
\newblock Reinforced latent reasoning for llm-based recommendation.
\newblock \emph{arXiv preprint arXiv:2505.19092}, 2025{\natexlab{c}}.

\bibitem[Zhang et~al.(2025{\natexlab{d}})Zhang, Zheng, Wu, Zhang, Lin, Yu, Liu, Zhou, and Lin]{zhang2025lessonsdevelopingprocessreward}
Zhenru Zhang, Chujie Zheng, Yangzhen Wu, Beichen Zhang, Runji Lin, Bowen Yu, Dayiheng Liu, Jingren Zhou, and Junyang Lin.
\newblock The lessons of developing process reward models in mathematical reasoning, 2025{\natexlab{d}}.
\newblock \url{https://arxiv.org/abs/2501.07301}.

\bibitem[Zhao et~al.(2025)Zhao, Xu, and Li]{zhao2025reasontorecommendusinginteractionofthoughtreasoning}
Keyu Zhao, Fengli Xu, and Yong Li.
\newblock Reason-to-recommend: Using interaction-of-thought reasoning to enhance llm recommendation, 2025.
\newblock \url{https://arxiv.org/abs/2506.05069}.

\bibitem[Zheng et~al.(2025{\natexlab{a}})Zheng, Zhang, Zhang, Lin, Lu, Yu, Liu, Zhou, and Lin]{zheng2025processbenchidentifyingprocesserrors}
Chujie Zheng, Zhenru Zhang, Beichen Zhang, Runji Lin, Keming Lu, Bowen Yu, Dayiheng Liu, Jingren Zhou, and Junyang Lin.
\newblock Processbench: Identifying process errors in mathematical reasoning, 2025{\natexlab{a}}.
\newblock \url{https://arxiv.org/abs/2412.06559}.

\bibitem[Zheng et~al.(2025{\natexlab{b}})Zheng, Lou, Cao, Wen, Ji, Lin, Lu, Han, Zhang, and Sun]{zheng2025criticcotboostingreasoningabilities}
Xin Zheng, Jie Lou, Boxi Cao, Xueru Wen, Yuqiu Ji, Hongyu Lin, Yaojie Lu, Xianpei Han, Debing Zhang, and Le~Sun.
\newblock Critic-cot: Boosting the reasoning abilities of large language model via chain-of-thoughts critic, 2025{\natexlab{b}}.
\newblock \url{https://arxiv.org/abs/2408.16326}.

\bibitem[Zhou et~al.(2022)Zhou, Liu, Zhai, Jiang, Gao, and Ji]{zhou2022prototype}
Xiong Zhou, Xianming Liu, Deming Zhai, Junjun Jiang, Xin Gao, and Xiangyang Ji.
\newblock Prototype-anchored learning for learning with imperfect annotations.
\newblock In \emph{ICML}, pages 27245--27267. PMLR, 2022.

\bibitem[Zhou et~al.(2024)Zhou, Tao, Zhu, Luo, Wang, and Han]{zhou2024can}
Zhanke Zhou, Rong Tao, Jianing Zhu, Yiwen Luo, Zengmao Wang, and Bo~Han.
\newblock Can language models perform robust reasoning in chain-of-thought prompting with noisy rationales?
\newblock \emph{NeurIPS}, 37:\penalty0 123846--123910, 2024.

\bibitem[Zhu et~al.(2024)Zhu, Wu, Guo, Hong, and Li]{zhu2024collaborative}
Yaochen Zhu, Liang Wu, Qi~Guo, Liangjie Hong, and Jundong Li.
\newblock Collaborative large language model for recommender systems.
\newblock In \emph{WWW}, pages 3162--3172, 2024.

\end{thebibliography}

\end{document}